\documentclass[letterpaper,twocolumn,10pt]{article}
\usepackage{usenix-2020-09}

\usepackage{multirow}
\usepackage[ruled,vlined,linesnumbered]{algorithm2e}
\usepackage[final]{changes}
\usepackage{tikz}
\usepackage{amsmath}
\usepackage{soul}
\usepackage{xcolor}
\usepackage{listings}
\usepackage{pythontex} 
\usepackage{subcaption}
\usepackage{comment}
\usepackage{adjustbox}
\usepackage{commath}
\usepackage{pythonhighlight}

\newcommand{\egg}[1] {}

\newcommand{\ourcompany}{{Alibaba}}

\newcommand{\eg}{\textit{e.g.}}
\newcommand{\ie}{\textit{i.e.}}
\newcommand{\etc}{\textit{etc.}}

\newcommand{\whale} {\textcolor{black}{Whale}}
\newcommand{\splitp} {\textcolor{black}{tensor model parallelism}}
\newcommand{\pipe} {\textcolor{black}{pipeline parallelism}}
\newcommand{\datap} {\textcolor{black}{data parallelism}}
\newcommand{\taskgraph} {\textit{TaskGraph}}
\newcommand{\ddp} {\textit{DP}}
\newcommand{\dmp} {\textit{MP}}
\newcounter{example}[section]

\usepackage{amsmath}

\usepackage{caption} 
\lstdefinestyle{Python}{
    language        = Python,
    keywordstyle    = \color{blue},
    commentstyle    = \color{gray}\ttfamily,
    deletekeywords={},
    otherkeywords={wh,with wh.cluster,wh.split,wh.replicate,wh.init,wh.pipeline,whale,wh.Config,pipeline.num_micro_batch,pipeline.num_stages,wh.set_default_strategy,auto.auto_parallel},
    frame=tblr,
    captionpos=b,                    
    escapeinside={(*}{*)},
    columns=fixed ,
     breaklines=true, 
}

\usepackage{filecontents}

\begin{document}

\date{}

\title{\whale{}: Efficient Giant Model Training over Heterogeneous GPUs}

\author{Xianyan Jia$^1$, Le Jiang$^1$, Ang Wang$^1$, Wencong Xiao$^1$, Ziji Shi$^{12}$, Jie Zhang$^1$, \\Xinyuan Li$^1$, Langshi Chen$^1$, Yong Li$^1$, Zhen Zheng$^1$, Xiaoyong Liu$^1$, Wei Lin$^1$\\
\\
$^1$Alibaba Group  $^2$National University of Singapore\\
}


\maketitle

\begin{abstract}
\textit{
The scaling up of deep neural networks has been demonstrated to be effective in improving model quality,
but also encompasses several training challenges in terms of training efficiency, programmability, and resource adaptability.
We present \whale{}, a general and efficient distributed training framework for giant models.
To support various parallel strategies and their hybrids, \whale{} generalizes the programming interface by defining two new primitives in the form of model annotations, allowing for incorporating user hints.
The \whale{} runtime utilizes those annotations and performs graph optimizations to transform a local deep learning DAG graph for distributed multi-GPU execution.
\whale{} further introduces a novel hardware-aware parallel strategy, which improves the performance of model training on heterogeneous GPUs in a balanced manner.
Deployed in a production cluster with 512 GPUs, 
\whale{} successfully trains an industry-scale multimodal model with over ten trillion model parameters, named M6, 
demonstrating  great scalability and efficiency.
}
\end{abstract}
\section{Introduction}
\label{sec:intro}
The training of large-scale deep learning (DL) models has been extensively adopted in various fields, including computer vision\cite{dosovitskiy2020image,liu2021swin}, natural language understanding\cite{vaswani2017attention,brown2020language,raffel2019exploring,shoeybi2019megatron}, machine translation\cite{lepikhin2020gshard,fedus2021switch}, and others. 
The scale of the model parameters increases from millions to trillions, which significantly improves the model quality \cite{brown2020language,kaplan2020scaling};
but at the cost of considerable efforts to efficiently distribute the model across GPUs.
The commonly used \datap{} ($DP$) strategy is a poor fit, since it requires the model replicas in GPUs perform gradient synchronization proportional to the model parameter size for every mini-batch, 
thus easily becoming a bottleneck for giant models.
Moreover, training trillions of model parameters requires  terabytes of GPU memory  at the minimum, 
which is far beyond the capacity of a single GPU.

To address the aforementioned challenges, a series of new parallel strategies in training DL models have been proposed, including model parallelism (\dmp{})~\cite{krizhevsky2014one}, pipeline parallelism~\cite{huang2018gpipe, narayanan2019pipedream}, \etc{}
For example, differing from the \ddp{} approach where each GPU maintains a model replica, \dmp{} partitions model parameters into multiple GPUs, avoiding gradient synchronization but instead letting tensors flow across GPUs.

Despite such advancements, new parallel strategies also introduce additional challenges. 
First, different components of a model might require different parallel strategies.
Consider a large-scale image classification task with 100K classes, where the model is composed of $ResNet50$\cite{he2016deep} for feature extraction and Fully-Connected ($FC$) layer for classification.
The parameter size of $ResNet50$ is 90 MB, and the parameter size of $FC$ is 782 MB. If  $DP$ is applied to the whole model, the gradient synchronization of $FC$ will become the bottleneck. 
One better solution is to apply $DP$ to $ResNet50$ and apply $MP$ to $FC$ (Section~\ref{sec:opportunities}).
As a result, the synchronization overhead can be reduced by 89.7\%, thereby achieving better performance~\cite{krizhevsky2014one}.

Additionally, using those advanced parallel strategies increases user efforts significantly.
To apply $DP$ in distributed model training, model developers only need to program the model for one GPU and annotate a few lines,
and DL frameworks can replicate the execution plan among multiple GPUs automatically~\cite{pytorchddp2020}. 
However, adopting advanced parallelism strategies might make different GPUs process different partitions of the model execution plan,
which is difficult to achieve automatically and efficiently~\cite{jia2019soap, wang2019tofu}.
Therefore, significant efforts are required for users to manually 
place computation operators, coordinate pipeline among mini-batches, implement equivalent distributed operators,
 and control computation-communication overlapping, \etc{}~\cite{lepikhin2020gshard, rasley2020deepspeed, shoeybi2019megatron, shazeer2018mesh}.
Such an approach exposes low-level system abstractions and requires users to understand system  implementation details when programming the models,
which greatly increases the amount of user effort.

Further, the training of giant models requires huge computing resources. 
In industry, the scheduling of hundreds of homogeneous high-end GPUs usually requires a long queuing time. 
Meanwhile, heterogeneous GPUs can be obtained much easier
 (\eg, a mixture of P100\cite{nvidiap100} and V100\cite{nvidiav100})\cite{jeon2019analysis, weng2022mlasas}.
But training with heterogeneous GPUs efficiently is even more difficult,  since both the computing units and the memory capacity of GPUs need to be considered when building the model. 
In addition, due to the dynamic scheduling of GPUs, users are unaware of the hardware specification when building their models, 
which brings a gap between model development and the hardware environment.

We propose \whale{}, 
a deep learning framework designed for training giant models.
Unlike the aforementioned approaches in which  the efficient model partitions are searched  automatically
or low-level system abstractions and implementation details are exposed to users,
we argue that deep learning frameworks should offer high-level abstractions properly to support complicated parallel strategies by utilizing user hints, 
especially when considering the usage of heterogeneous GPU resources.
Guided by this principle, \whale{} strikes a balance by extending two necessary primitives on top of TensorFlow~\cite{abadi2016tensorflow}.
Through annotating a local DL model with those primitives,
\whale{} supports all existing parallel strategies and their combinations, which is achieved by automatically rewriting the deep learning execution graph.
This design choice decouples the parallel strategies from model code, and lowers them into dataflow graphs,
 which not only reduces user efforts but also enables graph optimizations and resources-aware optimizations for efficiency and scalability.
In this way, \whale{} eases users from the complicated execution details of giant model training, 
such as scheduling parallel executions on multiple devices, and balancing computation workload among heterogeneous GPUs.
Moreover, \whale{} introduces a hardware-aware load balancing algorithm when generating a distributed execution plan, 
which bridges the gap between model development and the heterogeneous runtime environment.

We summarize the key contributions of \whale{} as follows: 
\begin{enumerate}
  \item For carefully balancing user efforts and distributed graph optimization requirements, 
  \whale{} introduces two new high-level primitives to express all existing parallel strategies as well as their hybrids.
  \item By utilizing the annotations for graph optimization, \whale{} can transform  local models into  distributed models, and train them on multiple GPUs efficiently and automatically.
  \item \whale{} proposes a hardware-aware load balancing algorithm, which is seamlessly integrated with parallel strategies to accelerate training on heterogeneous GPUs.
  \item \whale{} demonstrates its capabilities by setting a new milestone in training the largest multi-modality pretrained model M6~\cite{lin2021m6} with ten trillion model parameters,
   which requires only four lines of code change to scale the model and run on 512 NVIDIA V100M32 GPUs (Section~\ref{sec:exp-m6}).
\end{enumerate}

\whale{} has been deployed as a production system for large-scale deep learning training at \ourcompany{}. 
Using heterogeneous GPUs, further speedup of Bert-Large\cite{devlin2018bert}, Resnet50\cite{he2016deep}, and GNMT\cite{wu2016google}
 from 1.2x to 1.4x can be achieved owing  to the hardware-aware load balancing algorithm in \whale{}.
\whale{} also demonstrates its capabilities in the training of industry-scale models.
With only four-line changes to a local model, \whale{} can train a Multi-Modality to Multi-Modality Multitask Mega-transformer model with 10 billion parameters (M6-10B)
on 256 NVIDIA V100 GPUs (32GB), achieving 91\% throughput in scalability.
What's more, \whale{} scales to ten trillion parameters in model training of M6-10T using tensor model parallelism on 512 V100 GPUs (32GB), setting a new milestone in large-scale deep learning model training.
\section{Background and Motivation}
\label{sec:motivation}



\begin{figure*}[t]
	\begin{minipage}[t]{0.24\textwidth} %
		\centering 
		\includegraphics[width=0.95\linewidth]{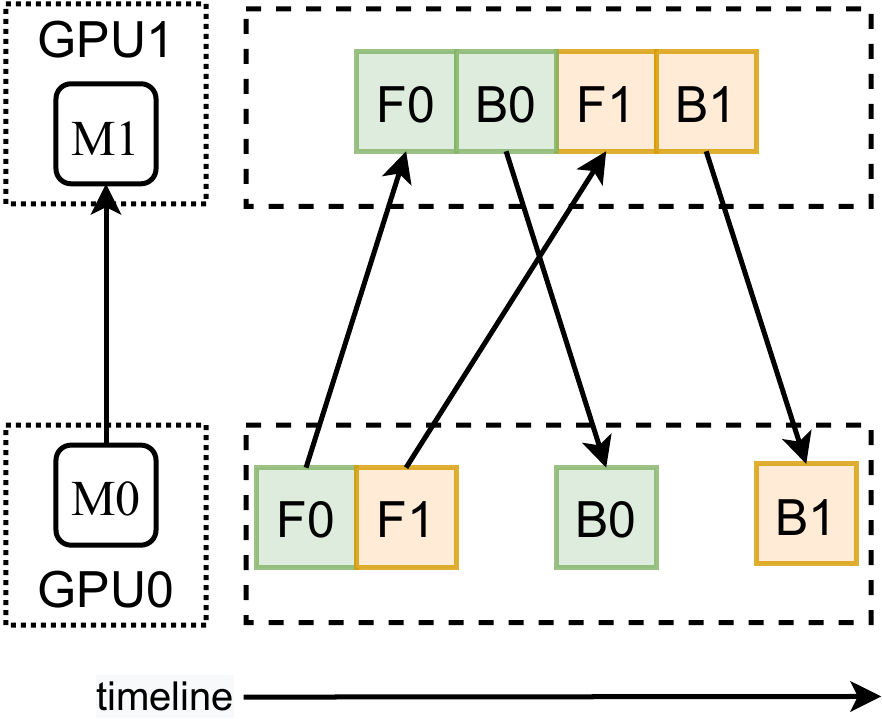} 
		\vspace{-0.1in}
		\caption{Pipeline parallelism of 2 micro-batches on 2 GPUs.}
		\label{fig:pipeline-parallel} 
	\end{minipage}
	\hfill
	\begin{minipage}[t]{0.24\textwidth}
		\centering 
		\includegraphics[width=0.95\linewidth]{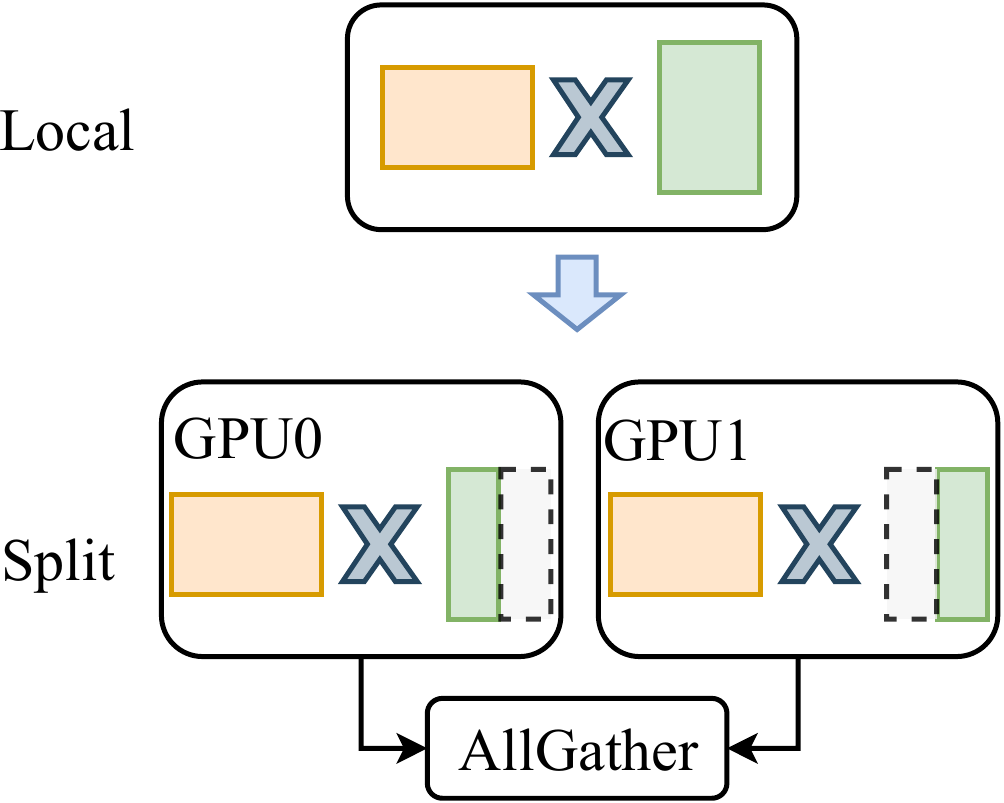} 
		\vspace{-0.1in}
		\caption{Tensor model parallelism for \textit{matmul} on 2 GPUs.}
		\label{fig:tensor-model-parallel} 
	\end{minipage}
	\hfill
	\begin{minipage}[t]{0.24\textwidth}
		\centering 
		\includegraphics[width=0.95\linewidth]{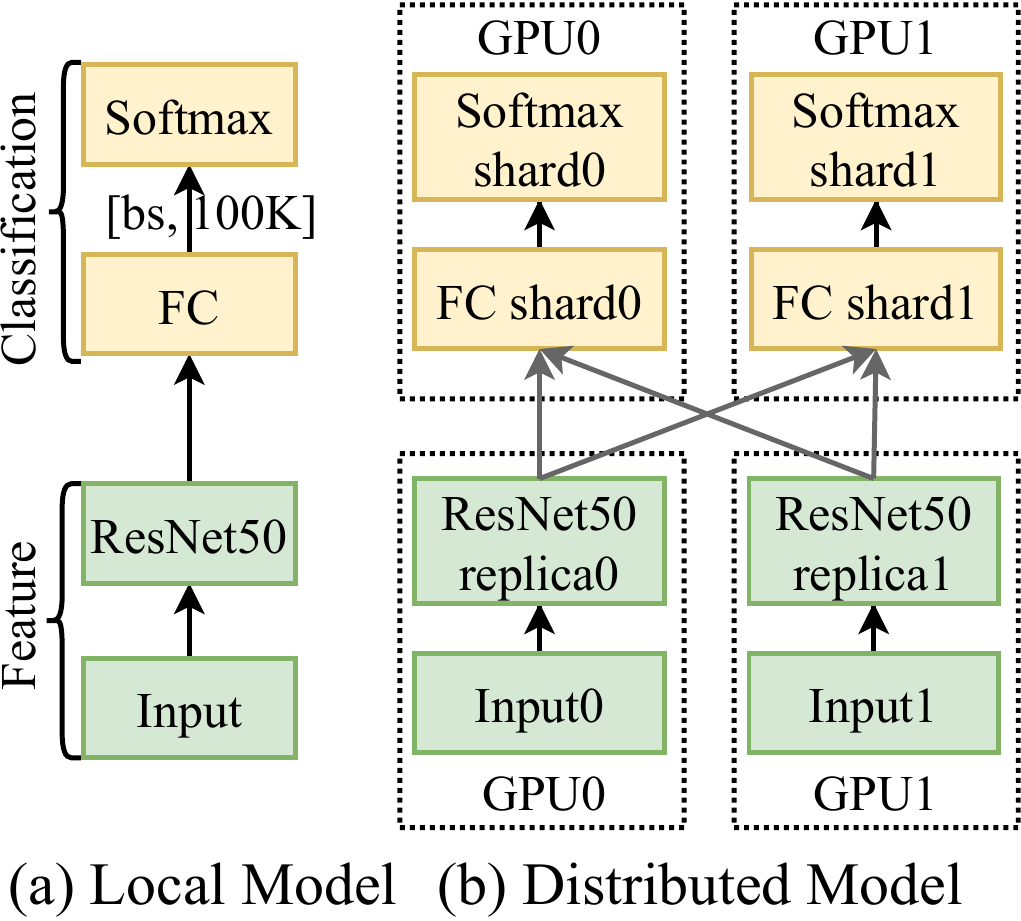} 
		\vspace{-0.1in}
		\caption{Hybrid parallelism for image classification.} 
		\label{fig:hybrid-parallel} 
	\end{minipage}%
	\hfill
	\begin{minipage}[t]{0.24\textwidth} 
		\centering 
		\includegraphics[width=0.95\linewidth]{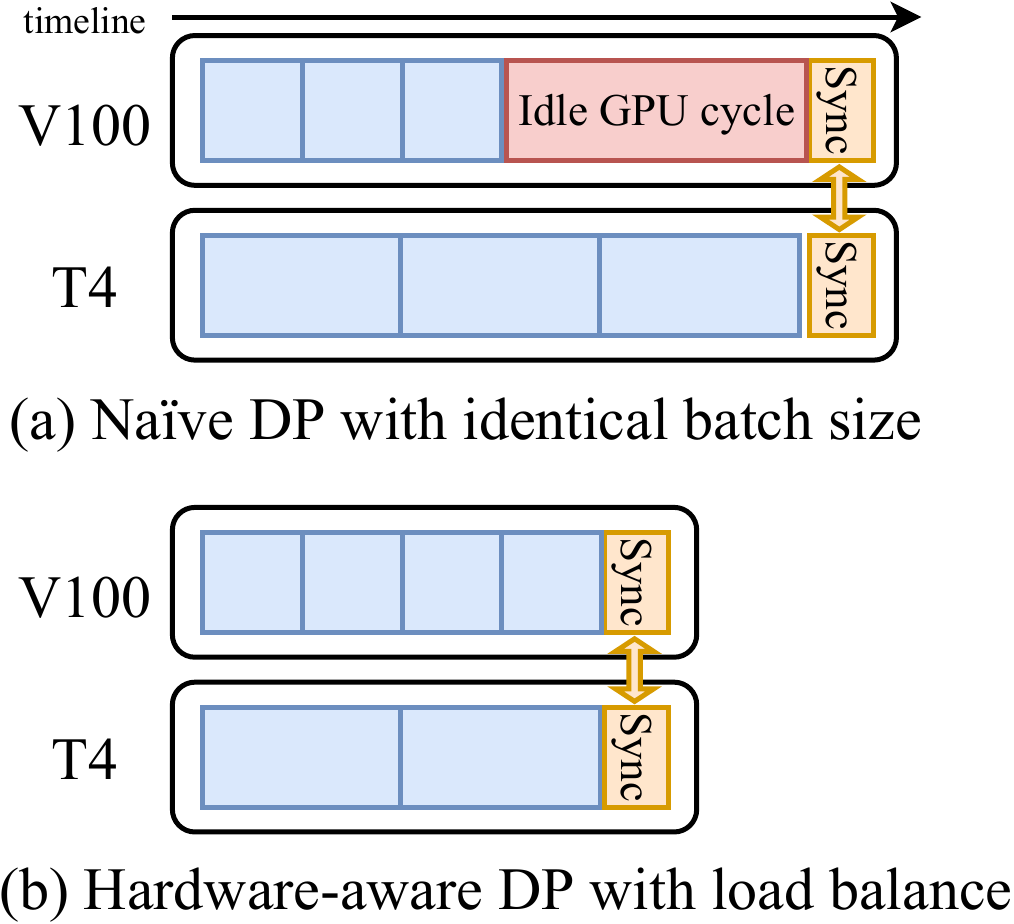} 
		\vspace{-0.1in}
		\caption{Data parallelism on heterogeneous GPUs} 
		\label{fig:dp-heter} 
	\end{minipage} 
	\vspace{-0.2in}
\end{figure*}

In this section, we first recap the background of distributed DL model training, especially the parallel strategies for large model training.
We then present the importance and the challenges of utilizing heterogeneous GPU resources.
Finally, we discuss the gaps and opportunities among existing approaches to motivate the design of a new training framework.

\subsection{Parallel Strategies}
\label{sec:strategy}

Deep learning training often consists of millions of iterations, referred to as \textit{mini-batches}.
A typical mini-batch includes several phases to process data for model updating. 
Firstly, the training data is fed into the model layer-by-layer to calculate a set of scores, known as a \textit{forward} pass. 
Secondly, a training loss is calculated between the produced scores and desired scores, which is then utilized to compute gradients for model parameters, referred to as a \textit{backward} pass. 
Finally, the gradients scaled by a learning rate are used to update the model parameters and optimizer states.

\paragraph{Data parallelism.}
Scaling to multiple GPUs, data parallelism is a commonly adopted strategy where each worker holds a full model replica to process different training data independently.
During the backward pass of every mini-batch, the gradients are averaged through worker synchronization.
Therefore, the amount of communication is proportional to the model parameter size.

\paragraph{Pipeline Parallelism.}
As shown in Figure~\ref{fig:pipeline-parallel}, a DL model is partitioned into two modules, \ie, M0 and M1 (which are also named pipeline stages), which are placed on 2 GPUs respectively.
The training data of a mini-batch is split into two smaller \textit{micro-batches}.
In particular, GPU0 starts with the forward of the $1^{st}$ micro-batch on M0, and then it switches to process the forward of the $2^{nd}$ micro-batch
while sending the output of the $1^{st}$ micro-batch to GPU1.
After GPU1 finishes processing forward and backward of the $1^{st}$ micro-batch on M1,
GPU0 continues to calculate the backward pass for M0 after receiving the backward output of M1 from GPU1. 
Therefore, micro-batches are pipelined among GPUs, which requires the runtime system to balance the load and overlap computation and communication carefully~\cite{huang2018gpipe,yang2021pipemare,narayanan2019pipedream,fan2020dapple}.
The model parallelism~\cite{distbelief2012, adam2014} can be treated as a special case of pipeline parallelism with only one micro-batch.

\paragraph{Tensor Model Parallelism.}
With the growing model size, to process DL operators beyond the  memory capacity of the GPU,
or to avoid significant communication overhead across model replicas, 
an operator (or several operators) might be split over multiple GPUs.
The tensor model parallelism strategy partitions the input/output tensors
and requires an equivalent distributed implementation for the corresponding operator. 
For example, Figure~\ref{fig:tensor-model-parallel} illustrates 
the tensor model parallelism strategy for a \textit{matmul} operator (\ie, matrix multiplication) using 2 GPUs.
A matmul operator can be replaced by two matmul operators, wherein each operator is responsible for half of the original computation. 
An extra all-gather operation is required to merge the distributed results.



In selecting a proper parallel strategy for model training,  both model properties and resources need to be considered.
For example, transformer~\cite{vaswani2017attention} is an important model in natural language understanding, 
which can be trained efficiently using pipeline parallelism on a few GPUs (\eg, 8 V100 GPUs with NVLINK~\cite{nvlink}).
However, pipeline parallelism does not scale well with more GPUs (\eg, 64 V100 GPUs).
Given more GPUs, each training worker is allocated with fewer operators, 
of which the GPU computation is not sufficient enough to overlap with the inter-worker communication cost, resulting in poor performance. 
Therefore, a better solution is to apply hybrid parallelism, 
where model partitions can be applied with different parallel strategies in combination, 
and parallel strategies can be nested.
Particularly, for the training of a transformer model on 64 GPUs,  the model parameters can be partitioned into 8 GPUs using a pipeline strategy,
and apply model replica synchronization among 8 pipelined groups using nested data parallelism.
 Moreover, different parallel strategies can also apply to different model partitions for a hybrid. 
As an example, a large-scale image classification model (\ie, 100K categories) consists of the image feature extraction partition and the classification partition.
The image feature extraction partition requires a significant amount of computation on fewer model parameters.
Conversely, the classification partition includes low-computation \textit{fully-connected} and \textit{softmax} layers,
which are often 10x larger in model size compared to that of image feature extraction.
Therefore, adopting a homogeneous parallel strategy will hinder the performance of either partitions.
Figure~\ref{fig:hybrid-parallel} illustrates a better hybrid parallelism approach,
in which data parallelism is applied for features extraction partition,  tensor model parallelism is adopted for classification partition,
and the two are connected.

\subsection{Heterogeneity in GPU Clusters}
\label{sec:bg-heto}

Training a giant model is considerably resource-intensive~\cite{narayanan2021efficient, fedus2021switch}. 
Moreover, distributed model training often requires resources to arrive at the same time (\ie, gang schedule~\cite{jeon2019analysis, antman2020}).
In industry, the shared cluster for giant model training is usually mixed with various types of GPUs (\eg, V100, P100, and T4)
for both model training and inference~\cite{weng2022mlasas}.
Training giant models over heterogeneous GPUs lowers the difficulty of collecting all required GPUs (\eg, hundreds or thousands of GPUs) simultaneously,
therefore speeding up the model exploration and experiments.
However,  deep learning frameworks encounter challenges in efficiently utilizing heterogeneous resources.
Different types of GPUs are different in terms of GPU memory capacity (\eg, 16GB for P100 and 32GB for V100) and GPU computing capability,
which natively introduces an imbalance in computational graph partition and deep learning operator allocation.
Figure~\ref{fig:dp-heter} illustrates  training a model using data parallelism on two heterogeneous GPUs, \ie, V100 and T4.
The V100 training worker completes forward and backward faster when training samples are allocated evenly,
thereby leaving idle GPU cycles before gradient synchronization at the end of every mini-batch.
Through the awareness of hardware when dynamically generating an execution plan, \whale{} allocates more training samples (\ie, batch-size=4) for V100 and the rest of 2 samples for T4
to eliminate the idle waiting time.
Combined with advanced parallel strategies and the hybrids over heterogeneous GPUs, 
 different GPU memory capacities and capabilities need to be further considered when partitioning the model for efficient overlapping, 
which is a complex process (Section~\ref{sec:hardware-aware}).
Model developers can hardly  consider all  resources issues  when programming, and we argue that developers should not have to. 
A better approach for a general deep learning framework would be  automatically generating the execution plan for heterogeneous resources adaptively.

\subsection{Gaps and Opportunities}
\label{sec:opportunities}

Recent approaches~\cite{rasley2020deepspeed, shoeybi2019megatron, shazeer2018mesh,huang2018gpipe,lepikhin2020gshard} have been proposed for giant model training,
however, with limitations as a general DL framework.
Firstly, they only support a small number of parallel strategies, 
which lack a unified abstraction to support all of the parallel strategies and the hybrids thereof.
Secondly, significant efforts are required in code modifications to utilize the advanced parallel strategies, 
compared with local model training and \ddp{} approach.
Mesh-tensorflow~\cite{shazeer2018mesh} requires the re-implementation of DL operators in a distributed manner.
Megatron~\cite{shoeybi2019megatron}, GPipe~\cite{huang2018gpipe}, DeepSpeed~\cite{rasley2020deepspeed}, and GShard~\cite{lepikhin2020gshard}
require user code refactoring using the exposed low-level system primitives or
a deep understanding for the implementation of parallel strategies.
Thirdly, automatically parallel strategy searching is time-consuming for giant models.
Although Tofu~\cite{wang2019tofu} and SOAP~\cite{jia2019soap} accomplish  model partitioning and replication automatically 
through computational graph analysis,
 the search-based graph optimization approach has high computational complexity, 
which is further positively associated with 
the number of model operators (\eg, hundreds of thousands of operators for GPT3\cite{brown2020language}) and allocated GPUs (\eg, hundreds or thousands),
making such an approach impractical when applying to giant model training.
Finally, due to the heterogeneity in both GPU computing capability and memory, 
parallel strategies should be used adaptively and dynamically.

There are significant gaps in supporting giant model training using existing DL frameworks.
Exposing low-level interfaces dramatically increases user burden and limits system optimization opportunities.
Users need to understand the implementation details of distributed operators and handle the overlapping of computation with communication,
which is hard for model developers.
Using a low-level approach tightly couples model code to a specific parallel strategy, 
which requires code rewriting completely when switching between parallel strategies (\ie, from pipeline parallelism to tensor model parallelism).
More constraints are introduced to model algorithm innovations, because the efforts of implementing a new module correctly in hybrid strategies are not trivial,
let alone consider the performance factors such as load balancing and overlapping. 
From the system aspect, seeking a better parallel strategy or a combination using existing ones also
requires rewriting user code, demanding a deep understanding of the DL model.

To address the aforementioned challenges, \whale{} explores a new approach that supports various parallel strategies
 while minimizing user code modifications.
By introducing new unified primitives, users can focus on implementing the model algorithm itself, 
while switching among various parallel strategies by simply changing the annotations. 
\whale{} runtime utilizes the user annotations as hints to select parallel strategies at best effort
with automatic graph optimization under a limited search scope.
 \whale{} further considers heterogeneous hardware capabilities using a balanced algorithm, making resource heterogeneity transparent to users.

\section{Design}
\label{sec:parallel}

In this section, we first introduce key abstractions and parallel primitives which can express flexible parallelism strategies with easy programming API (Section~\ref{sec:abstraction}).
Then, we describe our parallel planner that transforms a local model with parallel primitives into a distributed model,
 through partitioning \taskgraph{}s, inserting bridge layers to connect hybrid strategies, and placing \taskgraph{}s on distributed devices (Section~\ref{sec:plan}).
In the end, we propose a hardware-aware load balance algorithm to speed up the training with heterogeneous GPU clusters (Section~\ref{sec:hardware-aware}).

\subsection{Abstraction}
\label{sec:abstraction}

\subsubsection{Internal Key Concepts}
Deep learning frameworks such as TensorFlow\cite{abadi2016tensorflow} provide low-level APIs for distributed computing,
 but is short of abstractions to represent advanced parallel strategies such as pipeline.
 The lack of proper abstractions makes it challenging in the understanding and implementation of complicated strategies in a unified way.
 Additionally, placing model operations to physical devices properly is challenging for complicated hybrid parallel strategies, especially in heterogeneous GPU clusters.
 \whale{} introduces two internal key concepts, i.e., \textit{\taskgraph{}} and \textit{VirtualDevice}.
 \textit{\taskgraph{}} is used to modularize operations for applying a parallel strategy.
\textit{VirtualDevice} hides the complexity of mapping operations to physical devices.
 The two concepts are abstractions of internal system design and are not exposed to users.



\textbf{\textit{\taskgraph{}}}(\textit{TG}) is a subset of the model for parallel transformation and execution.
One model can have one or more non-overlapping \taskgraph{}s.
We can apply parallel strategies to each \taskgraph{}.
By modularizing model operations into \taskgraph{}s, \whale{} can apply different strategies to different model parts,
 as well as scheduling the execution of \taskgraph{}s in a pipeline.
A \taskgraph{} can be further replicated or partitioned.
For example, in \datap{}, the whole model is a \taskgraph{}, which can be replicated to multiple devices.
In \pipe{}, one pipeline stage is a \taskgraph{}.
In \splitp{}, we can shard the \taskgraph{} into multiple submodules for parallelism.



\textbf{\textit{VirtualDevice}} (VD) is the logical representation of computing resources,
with one \textit{VirtualDevice} having one or more physical devices.
\textit{VirtualDevice} hides the complexity of device topology, computing capacity as well as  device placement from users.
One \textit{VirtualDevice} is assigned to one \taskgraph{}.
Different \textit{VirtualDevice}s are allowed to have different or the same physical devices.
For example, VD0 contains physical devices GPU0 and GPU1,
VD1 contains physical devices GPU2 and GPU3 (different from VD0),
and VD2 contains physical devices GPU0 and GPU1 (the same as VD0).

\subsubsection{Parallel Primitives}
\label{sec:primitive}
The parallel primitive is a Python context manager,
 where operations defined under it are modularized as one \taskgraph{}.
 Each parallel primitive has to be configured with a parameter $device\_count$,
  which is used to generate a \textit{VirtualDevice} by mapping the $device\_count$ number of physical devices. 
  \whale{} allows users to suggest parallel strategies with two unified primitives, \ie, $replicate$ and $split$.
The two primitives can express all existing parallel strategies, as well as a hybrid of them\cite{huang2018gpipe,narayanan2019pipedream,shoeybi2019megatron,lepikhin2020gshard,krizhevsky2014one}.

\textbf{\textit{replicate(device\_count)}} annotates a \taskgraph{} to be replicated.
 $device\_count$ is the number of devices used to compute the \taskgraph{} replicas.
 If $device\_count$ is not set, \whale{} allocates a \taskgraph{} replica per device.
If a \taskgraph{} is annotated with $replicate(2)$, it is replicated to $2$ devices, with each \taskgraph{} replica consuming half of the mini-batch.
Thus the mini-batch size for one model replica is kept unchanged.

\textbf{\textit{split(device\_count)}} annotates a \taskgraph{} to apply intra-tensor sharding.
The $ device\_count $ denotes the number of partitions to be sharded.
Each sharded partition is placed on one device.
For example, $split(2)$ shards the \taskgraph{} into 2 partitions and placed on 2 devices respectively.





 \begin{figure}[t]
	\vspace{-0.03in}
	\begin{minipage}[t]{0.45\linewidth} %
		\begin{lstlisting}[style = Python, caption=Pipeline with 2 \taskgraph{}s, label=case:pipeline]
import whale as wh
wh.init(wh.Config({
  "num_micro_batch": 8}))
with wh.replicate(1):
  model_stage1()
with wh.replicate(1):
  model_stage2()
      \end{lstlisting}
	\end{minipage}
  	\hspace{0.1in}
  	\begin{minipage}[t]{0.55\linewidth} %
		\begin{lstlisting}[style = Python, caption=Hybrid of replicate and split, label=case:classification]
import whale as wh
wh.init()
with wh.replicate(total_gpu):
  features = ResNet50(inputs)
with wh.split(total_gpu):
  logits = FC(features)
  predictions = Softmax(logits)
      \end{lstlisting}
	\end{minipage}

	 \vspace{-0.2in}
\end{figure}

The  parallel primitives can be used in combination to apply different parallel strategies to different partitions of the model.
Additionally, \whale{} also provides JSON Config API to enable system optimizations.
The config  $auto\_parallel$ is used to enable automatic \taskgraph{} partitioning given a provided partition number $num\_task\_graph$,
 which further eases the programming for users and is necessary for hardware-aware optimization when resource allocation is dynamic (Section~\ref{sec:hardware-aware}).
In \whale{}, pipeline parallelism is viewed as an efficient inter-\taskgraph{} execution strategy.
\whale{} uses the config $num\_micro\_batch$ to enable efficient pipeline parallelism among \taskgraph{}s  when the value is greater than 1.
In this way, \whale{} decouples the generation of \taskgraph{} from the choice of pipeline parallelism strategies\cite{narayanan2019pipedream,fan2020dapple,huang2018gpipe}.
The system can easily extend to incorporate more pipeline strategies (\eg, swap the execution order of $B0$ and $F1$ for M1 in Figure~\ref{fig:pipeline-parallel}).

Besides the combination of parallel strategies or pipeline parallelism, \whale{} further supports nested data parallelism to the whole parallelized model.
Nested data parallelism is enabled automatically when the number of available devices is times of total devices requested by \taskgraph{}s.

Example~\ref{case:pipeline} shows an example of pipeline parallelism with two \taskgraph{}s, with each \taskgraph{} being configured with 1 device.
The pipeline parallelism is enabled by configuring the $pipeline.num\_micro\_batch$ to 8.
The total device number of the two \taskgraph{}s is summed to 2.
If the available device number is 8, which is 4 times of total device number, \whale{} will apply a nested 4-degree data parallelism beyond the pipeline.
In contrast, when using two available devices, it is a pure pipeline.
Example~\ref{case:classification} shows a hybrid strategy that replicates $ResNet50$ feature part while splitting the $classification$ model part for the example in Figure~\ref{fig:hybrid-parallel}.

\begin{minipage}{0.9\linewidth}
\begin{lstlisting}[style = Python, caption=Auto pipeline, label=case:bert_pipeline_auto]
wh.init(wh.Config({"num_task_graph":2,
  "num_micro_batch":4,"auto_parallel":True}))
model_def()
\end{lstlisting}
\end{minipage}

Example~\ref{case:bert_pipeline_auto} shows an automatic pipeline example with two \taskgraph{}s.
When $auto\_parallel$ is enabled, \whale{} will partition the model into \taskgraph{}s automatically according to the computing resource capacity and the model structure. (Section~\ref{sec:hardware-aware})

\subsection{Parallel Planner}
\label{sec:plan}

\begin{figure}
  \centering
  \includegraphics[width=1\linewidth]{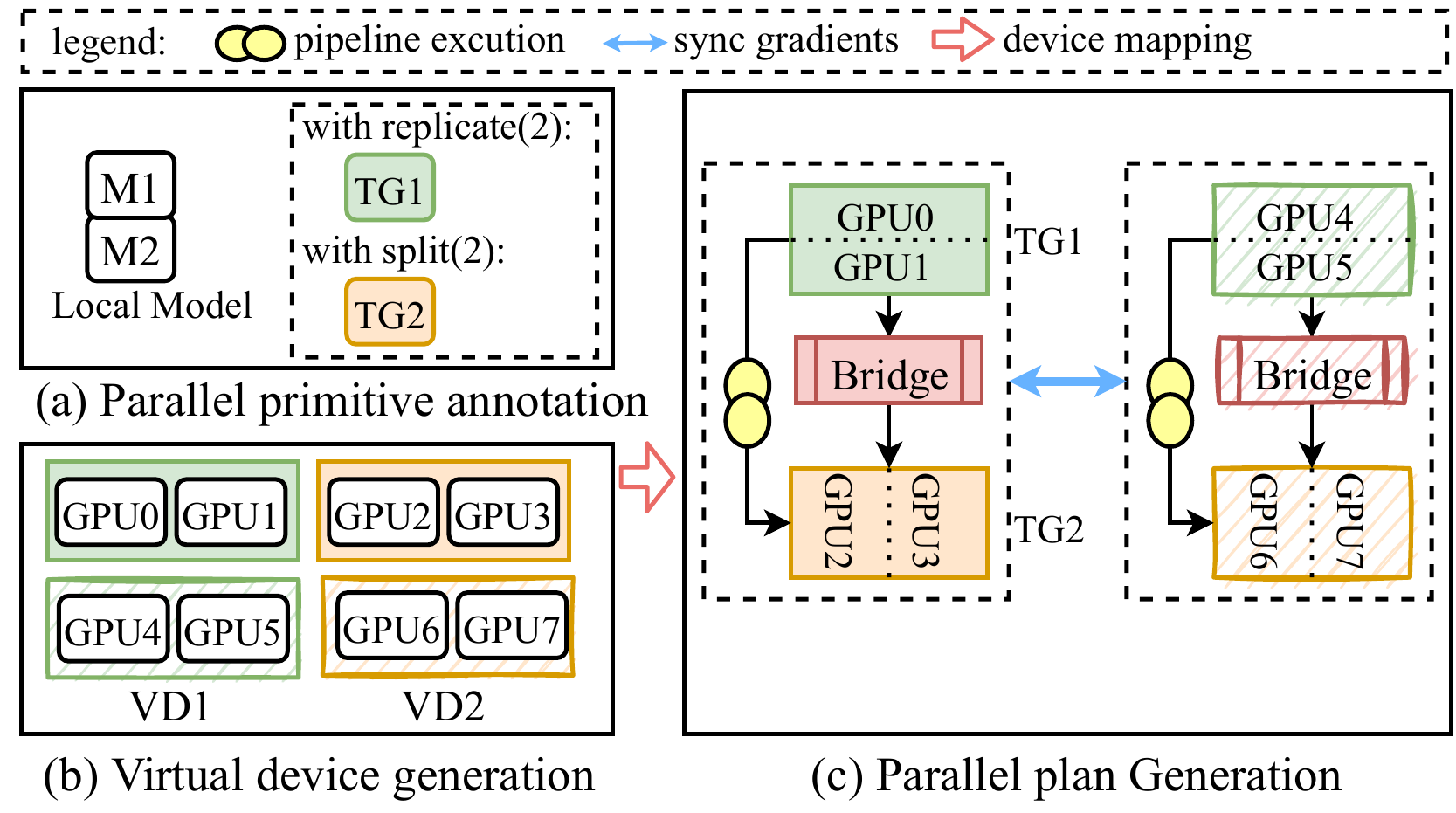}
		\vspace{-0.2in}
  \caption{\whale{} Overview}
  \label{fig:flow}
	\vspace{-0.2in}
\end{figure}
The parallel planner is responsible for producing an efficient parallel execution plan, which is the core of \whale{} runtime.
Figure~\ref{fig:flow} shows an overview of the parallel planner. The workflow can be described as follows: 
(a) The parallel planner takes a local model with optional user annotations, computing resources, and optional configs as inputs.
The model hyperparameters (\eg, batch size and learning rate), and computing resources (\eg, \#GPU and \#worker) are decided by the users manually.
While the parallel primitive annotations and configs (\eg, $num\_task\_graph$ and $num\_micro\_batch$) could be either be manual or decided by \whale{} automatically;
(b)  the \textit{VirtualDevice}s are generated  given computing resources and optional annotations automatically (Section~\ref{sec:cluster});
and (c)   the model is partitioned into \taskgraph{}s, and the \taskgraph{} is further partitioned  internally if $split$ is annotated.
Since we allow applying different strategies to different  \taskgraph{}s,
 there may exist an input/output mismatch among \taskgraph{}s. 
In such case, the planner will insert the corresponding bridge layer automatically between two \taskgraph{}s (Section~\ref{sec:bridge}).

\subsubsection{Virtual Device Generation}
\label{sec:cluster}
\textit{VirtualDevice}s are generated given the number of devices required by each \taskgraph{}.
Given $K$ requested physical devices $GPU_0, GPU_1, ..., GPU_K$ and a model with $N$ \taskgraph{}s, with corresponding device number $d_1, d_2, ... d_N$.
For the $i^{th}$ \taskgraph{}, \whale{} will generate a \textit{VirtualDevice} with $d_i$ number of physical devices.
The physical devices are taken sequentially for each \textit{VirtualDevice}.
As  mentioned in Section~\ref{sec:primitive},
 when the available device number $K$ is divisible by the total number of devices requested by all \taskgraph{}s $\sum_i^N{d_i}$,
  \whale{} will apply a nested \ddp{} of $\frac{K}{\sum_i^N{d_i}}$-degree  to the whole model.
In such case, we also replicate the corresponding \textit{VirtualDevice} for \taskgraph{} replica.
By default, devices are not shared among \taskgraph{}s. Sharing can be enabled to improve training
performance in certain model sharding cases by setting cluster configuration\footnote{\url{https://easyparallellibrary.readthedocs.io/en/latest/api/config.html\#clusterconfiguration}}.
\whale{} prefers to place one model replica (with one or more \taskgraph{}s) within a node, and replicates the model replicas across nodes.
 Advanced behaviors such as placing \taskgraph{} replicas within a node to utilize NVLINK for AllReduce communication can be achieved by setting the aforementioned configuration.
For example, as shown in Figure~\ref{fig:flow}, there are two \taskgraph{}s, and each \taskgraph{} requests 2 GPUs.
Two \textit{VirtualDevices} VD1 and VD2 are generated for two \taskgraph{}s.
VD1 contains $GPU0$ and $GPU1$, and VD2 contains $GPU2$ and $GPU3$.
As the number of available GPUs is 8, which is divisible by the total GPU number of \taskgraph{}s 4,
 a replica of \textit{VirtualDevices} can be generated but with different physical devices.

\subsubsection{TaskGraph Partitioning}
\label{sec:partition}
\whale{} first partitions a model into \taskgraph{}s, either by using explicit annotations or automatic system partitioning.
If a user annotation is given, operations defined within certain parallel primitive annotation compose a \taskgraph{}.
Otherwise, the system generates \taskgraph{}s based on the given config parameter $num\_task\_graph$ and hardware information.
The details of the hardware-aware model partitioning is described in Section~\ref{sec:hardware-aware}.

\begin{figure}
  \centering
  \includegraphics[width=0.8\linewidth]{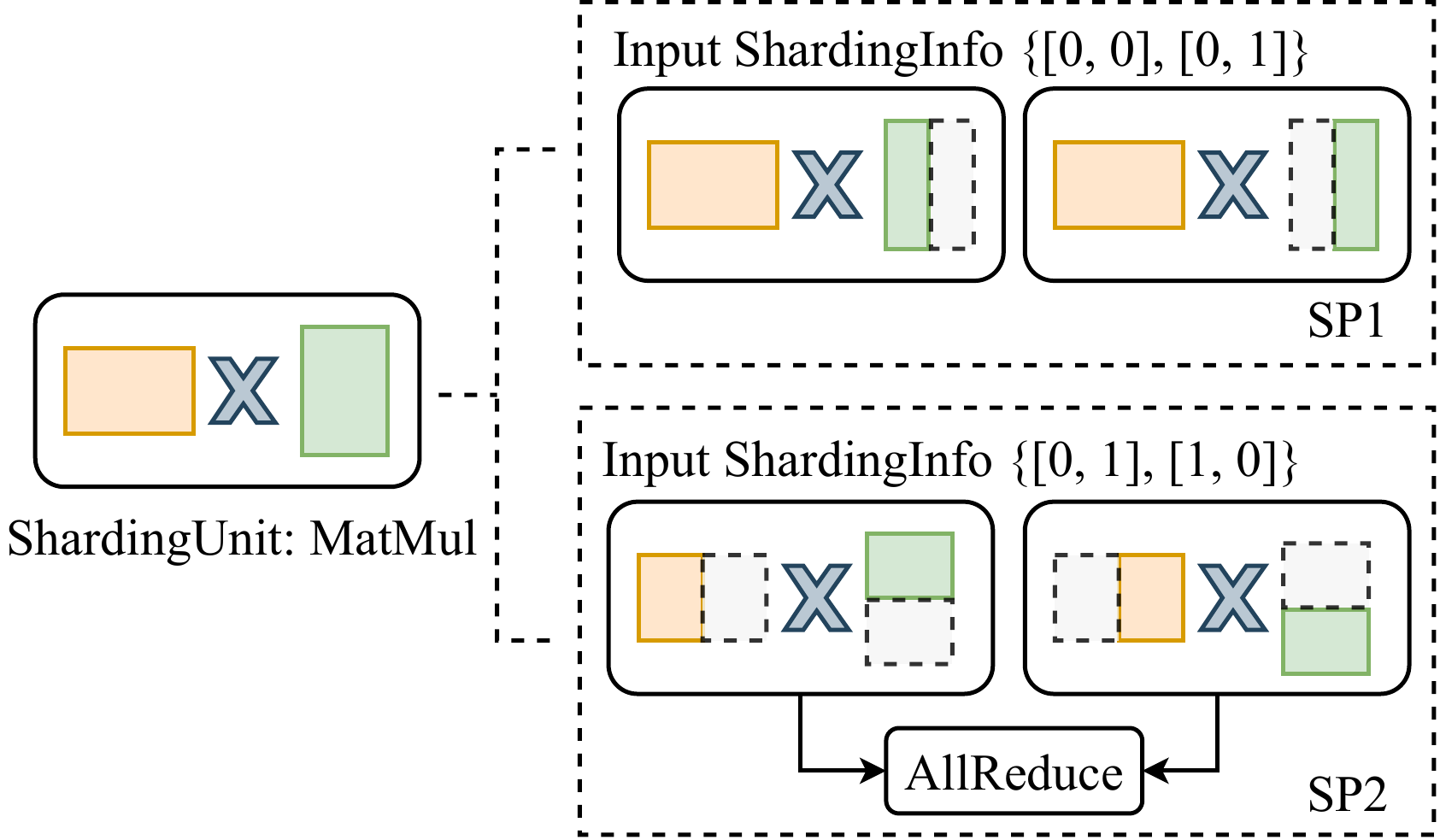}
  \vspace{-0.1in}
  \caption{Sharding pattern example for MatMul. One ShardingUnit can map to multiple sharding patterns.}
  \label{fig:sharding-pattern}
\end{figure}

If a \taskgraph{} is annotated with $split(k)$,
\whale{} will automatically partition it by matching and replacing sharding patterns  with a distributed implementation.
Before describing the sharding pattern, we introduce two terminologies for \splitp{}:
1) \textit{ShardingUnit} is a basic unit for sharding, and can be an operation or a layer with multiple operations;
and 2) \textit{ShardingInfo} is the tensor sharding information,
and is represented as a list $ [s_0, s_1,..., s_n] $ given a tensor with $n$ dimensions,
where $s_i$ represents whether to split the $i^{th}$ dimension,  1 means true and 0 means false.
For example, given a tensor with shape $ [6, 4] $, the \textit{ShardingInfo} $ [0, 1] $ indicates splitting in the second tensor dimension, whereas $[1, 1]$ indicates splitting in both dimensions.
A sharding pattern(SP) is a mapping from a \textit{ShardingUnit} and input \textit{ShardingInfo} to its distributed implementations.
For example, Figure~\ref{fig:sharding-pattern} shows two sharding patterns SP1 and SP2 with different input \textit{ShardingInfo} for \textit{ShardingUnit} MatMul.

To partition the \taskgraph{}, \whale{} first groups the operations in the split \taskgraph{} into multiple \textit{ShardingUnit}s by hooking TensorFlow ops API\footnote{TensorFlow Ops: \url{https://github.com/tensorflow/tensorflow/tree/r1.15/tensorflow/python/ops}}.
The \taskgraph{} sharding process starts by matching \textit{ShardingUnit}s to the predefined sharding patterns in a topology order.
A pattern is matched by a \textit{ShardingUnit} and input \textit{ShardingInfo}s.
If multiple patterns are matched,  the pattern with a smaller communication cost is selected.
\whale{} replaces the matched pattern of the original \textit{ShardingUnit} with its distributed implementation.

\subsubsection{Bridge Layer}
\label{sec:bridge}
When applying different parallel strategies to different \taskgraph{}s,
the input/output tensor number and shape may change due to different parallelism degrees or different parallel strategies,
 thereby resulting in a mismatch of input/output tensor shapes among \taskgraph{}s.
To address the mismatch, \whale{} proposes a \textit{bridge layer} to gather the distributed tensors and feed them to the next \taskgraph{}.

 

\begin{figure}[t]
  \centering
    \includegraphics[width=0.8\linewidth]{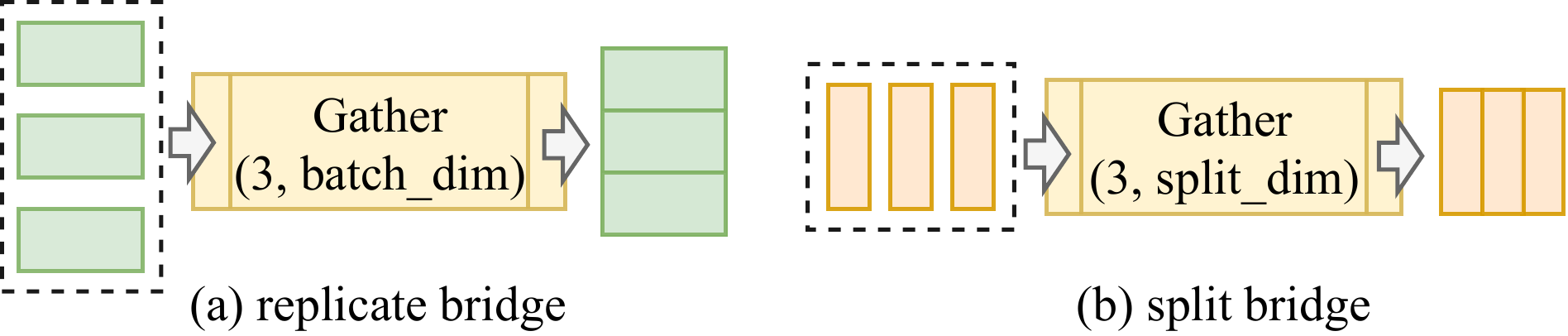}
		\vspace{-0.1in}
    \caption{Bridge patterns.}
    \label{fig:bridge}
    \vspace{-0.2in}
\end{figure}

\whale{} designs two \textit{bridge pattern}s for \textit{replicate} and \textit{split} respectively, as shown in Figure~\ref{fig:bridge}.
For \textit{replicate}, the \taskgraph{} is replicated to $N$ devices, with different input batches.
The \textit{bridge layer} gathers the outputs from different batches for concatenation in  batch dimension $batch\_dim$.
For \textit{split}, the outputs of \taskgraph{} are partitioned in split dimension $split\_dim$.
The \textit{bridge layer} gathers \taskgraph{} outputs for concatenation in $split\_dim$.
By using the \textit{bridge layer}, each \taskgraph{} can obtain a complete input tensor.
If the gather dimension of the \textit{bridge layer} is the same as the successor \taskgraph{} input partition dimension,
 \whale{} will optimize by fusing the aforementioned two operations to reduce the communication overhead.
As an example, if the outputs of the \taskgraph{} are gathered in the first dimension,
 and the inputs of the successor \taskgraph{} are partitioned in the same dimension,
then \whale{} will remove the above $gather$ and $partition$ operations.

\subsection{Hardware-aware Load Balance}
\label{sec:hardware-aware}

In this section, we describe how we utilize the hardware information to balance the workloads among \taskgraph{}s,
 which achieves high performance even in heterogeneous GPU clusters.
The \whale{} parallel planner obtains the hardware information from the cluster scheduler when the training job is launched,
and is responsible for both intra-\taskgraph{} and inter-\taskgraph{} load balancing.

\subsubsection{Intra-TaskGraph Load Balance} 
When the allocated devices are homogeneous, by default \whale{} distributes the workloads within a \taskgraph{} to multiple devices evenly.
However, when allocated with heterogeneous GPUs with different computing capacities (e.g., V100 and P100),
 the aforementioned identical distribution effectuates suboptimal performance.
Such performance can be attributed to a synchronization barrier at the end of \taskgraph{} execution, which leads to long idle GPU time for the faster GPU, as shown in Figure~\ref{fig:dp-heter}(a).
To improve the overall utilization of heterogeneous GPUs, we need to balance the computing according to the device's computing capacity.
The intra-\taskgraph{} load balance attempts to minimize the idle time within a \taskgraph{},
which is achieved by balancing the workloads proportional to device computing capacity while being subject to memory constraints.
For a \taskgraph{} annotated with $replicate$, \whale{} balances the workload by adjusting the batch size for each \taskgraph{} replica.
The local batch size on heterogeneous devices might differ due to the load balancing strategy (\whale{} keeps the global batch size unchanged). If batch-sensitive operators such as $BatchNorm$ exist, the local batch differences might have statistical effects.
Yet, no users suffer convergence issues when using heterogeneous training in \whale{}, which is probably due to the robustness of $DL$.
 Besides, techniques like SyncBatchNormaliazaion\footnote{\url{https://www.tensorflow.org/api_docs/python/tf/keras/layers/experimental/SyncBatchNor
 malization}} might help.
For a \taskgraph{} annotated with $split$, \whale{} balances the FLOP of a partitioned operation through uneven sharding in splitting dimension among multiple devices.

We profile the \taskgraph{} $TG$ on single-precision floating-point operations(FLOP) as $TG_{flop}$ and peak memory consumption as $TG_{mem}$.
Given $N$ GPUs, we collect the information for device $i$ including the single-precision FLOP per second as $DF_{i}$ and memory capacity as $DM_i$.
Assuming the partitioned load ratio on the device $i$ is $L_i$, we need to find a solution that minimizes the overall GPU waste, which is formulated in Formula~\ref{eq:balance}.
We try to minimize the ratio of the computational load of the actual model  for each device $L_i$ and the ratio of the computing capacity of the device over the total cluster computing capacity $DF_{i}/\sum_{i=0}^{N}{DF_{i}}$,
the maximum workload being bounded by the device memory capacity $DM_i$.

\begin{equation}
  \begin{aligned}
  \min          & \sum_{i}^{N}\norm{L_i - \frac{DF_{i}}{\sum_{i=0}^{N}{DF_{i}}}}\\
  \textrm{s.t.} & \sum_{i=0}^{N}L_i = 1;\  L_i * TG_{mem} <= DM_i, (i = 1,2, ..., N)  \\
  \end{aligned}
  \label{eq:balance}
\end{equation}

\begin{algorithm}[t]
  \small
  \caption{Memory-Constraint Load Balancing}	\label{algo:psvf}
  \SetKwProg{Function}{Function}{}{}
  \SetKwInput{Input}{Input}
  \Input{$TaskGraph~TG, VirtualDevice(N)$}
  $load\_ratios = \emptyset $; $mem\_utils = \emptyset$ ; $flop\_utils = \emptyset $ \\
  $oom\_devices = \emptyset$ ; $free\_devices = \emptyset $ \\
  \ForEach{$ i \in 0...N $\label{lst:init-start}} 
  {
    $ load\_ratios[i] = \frac{DF_{i}}{\sum_{i=0}^{N}{DF_{i}}}$ \\
    $ mem\_utils[i] = \frac{load\_ratios[i] * TG_{mem}}{DM_i} $ \\
    $ flop\_utils[i] = \frac{load\_ratios[i] * TG_{flop}}{DF_{i}} $ \\
    \If{$mem\_utils[i] > 1$}{$oom\_devices.append(i)$}
    \Else{$free\_devices.append(i)$\label{lst:init-end}}
  }
  \While{$oom\_devices \ne \emptyset ~\&~ free\_devices \ne \emptyset $\label{lst:psvf-start}}
  {
    $peak\_device = argmax(oom\_devices, key=mem\_utils)$ \\
    $valley\_device = argmin(free\_devices, key=(flop\_utils, mem\_utils))$ \\
    \If{$shift\_load(peak\_device, valley\_device) == success$}{$update\_profile(mem\_utils, flop\_utils)$\\$oom\_devices.pop(peak\_device)$}
    \Else{$free\_devices.pop(valley\_device)$\label{lst:psvf-end}}
  }
  \end{algorithm}
 
The load ratio $L_i$ in each device is initialized in proportional to the device's computing capacity, which ideally results in a most balanced partition.
However, when the memory constraint is not satisfied, we need to adjust the load allocation to avoid out-of-memory (OOM) errors,
while still trying to achieve good performance. 
  \whale{} proposes a memory-constraint balancing algorithm to balance the workloads among devices.
  The main idea of the algorithm is to shift the workload from the memory-overload device to a memory-free device with the lowest computation load.
  The details of the algorithm are illustrated in Algorithm~\ref{algo:psvf}.
  It takes a \taskgraph{} $TG$ and $VirtualDevice$ with $N$ physical devices as inputs.
  The algorithm  first initializes (line~\ref{lst:init-start}-\ref{lst:init-end}) the profiling results including
  1) $load\_ratios$ as the workload ratios of devices;
  2) $mem\_utils$ as the memory utilization of devices;
  3) $ flop\_utils $ as the FLOP utilization of devices;
  4) $oom\_devices$ records out of memory devices whose value in $mem\_utils$ is greater than 1;
  and 5) $free\_devices$ records devices that have free memory space.
  The algorithm then iteratively shifts the load from a memory-overload device to a memory-available device (line~\ref{lst:psvf-start}-\ref{lst:psvf-end}).
  It first finds a $peak\_device$ with maximum memory utilization from $oom\_devices$,
  then it finds a $valley\_device$ with available memory space and the lowest FLOP utilization.
  The $shift\_load$ function attempts to shift the workload from a $peak\_device$ to a $valley\_device$.
  For data parallelism, the batch size in the $peak\_device$ is decreased by $b$, and the batch size in the $valley\_device$ is increased by $b$.
  $b$ is the maximum number that the $valley\_device$ will not go $OOM$ after getting the load from the $peak\_device$.
  The profiling information for each device is updated after a successful workload shift is found.
  The aforementioned process iterates until the $oom\_devices$ are empty or the $free\_devices$ are empty.


\subsubsection{Inter-TaskGraph Load Balance}

\begin{figure}
  \centering
  \includegraphics[width=0.7\linewidth]{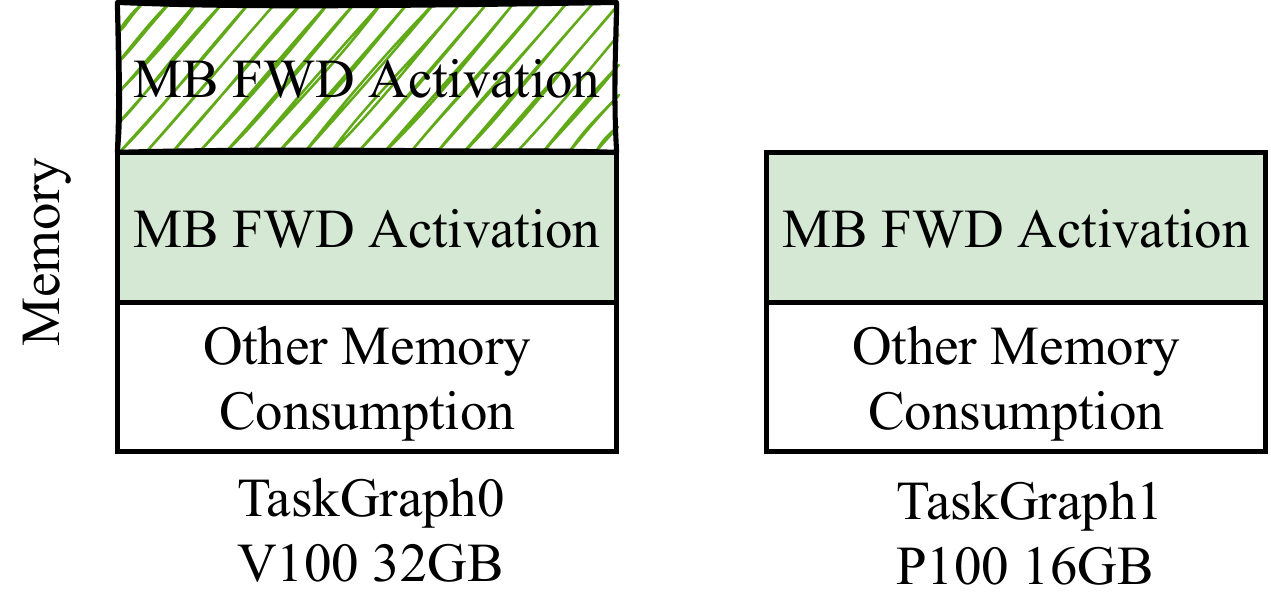}
  \vspace{-0.1in}
  \caption{Pipeline \taskgraph{}s on heterogeneous GPUs}
  \label{fig:pipeline-mem}
	\vspace{-0.2in}
\end{figure}

When multiple \taskgraph{}s are executed in a pipeline, we need to balance the inter-\taskgraph{} workloads on heterogeneous GPUs.
As we introduced in Section~\ref{sec:strategy}, pipeline parallelism achieves efficient execution by interleaving  forward/backward execution among multiple micro-batches.
For a model with $N$ \taskgraph{}s, the $i^{th}$ \taskgraph{} needs to cache $N-i$ forward activations\cite{narayanan2019pipedream}.
Notably, $i^{th}$ \taskgraph{} has to cache one more micro-batch forward activation than the previous \taskgraph{}.
Since activation memory is  proportional to batch size and often takes a large proportion of the peak memory, e.g., the activation memory VGG16 model with batch size 256 takes up around 74\% of the peak memory\cite{gao2020estimating},
resulting in uneven memory consumption among different \taskgraph{}s.
The different memory requirements of \taskgraph{}s motivate us to place earlier \taskgraph{}s on devices with higher memory capacity.
This can be achieved by sorting and reordering the devices in the corresponding \textit{VirtualDevice} by memory capacity, from higher to lower.
Figure~\ref{fig:pipeline-mem} shows the memory breakdown of the pipeline example (Figure~\ref{fig:pipeline-parallel}) with two \taskgraph{}s over heterogeneous GPUs V100 (32GB) and P100 (16GB), we prefer putting  TaskGraph0 to V100, which has a higher memory config. 
The \taskgraph{} placement heuristic is efficient for common Transformer-based
models (\ie, BertLarge and T5 in Figure~\ref{fig:heto-pipe}).
There might be cases where later stages contain large layers (\ie, large sparse embedding), which can be addressed in Algorithm~\ref{algo:psvf} on handling OOM errors.
After reordering the virtual device according to memory requirement,
we partition the model operations to \taskgraph{}s in a topological sort
and apply Algorithm~\ref{algo:psvf} to balance the computing FLOP among operations, subject to the memory bound of  the memory capacity of each device.

\begin{figure*}[t!]
  \begin{minipage}[t]{0.24\textwidth} %
    \centering
    \includegraphics[width=\textwidth]{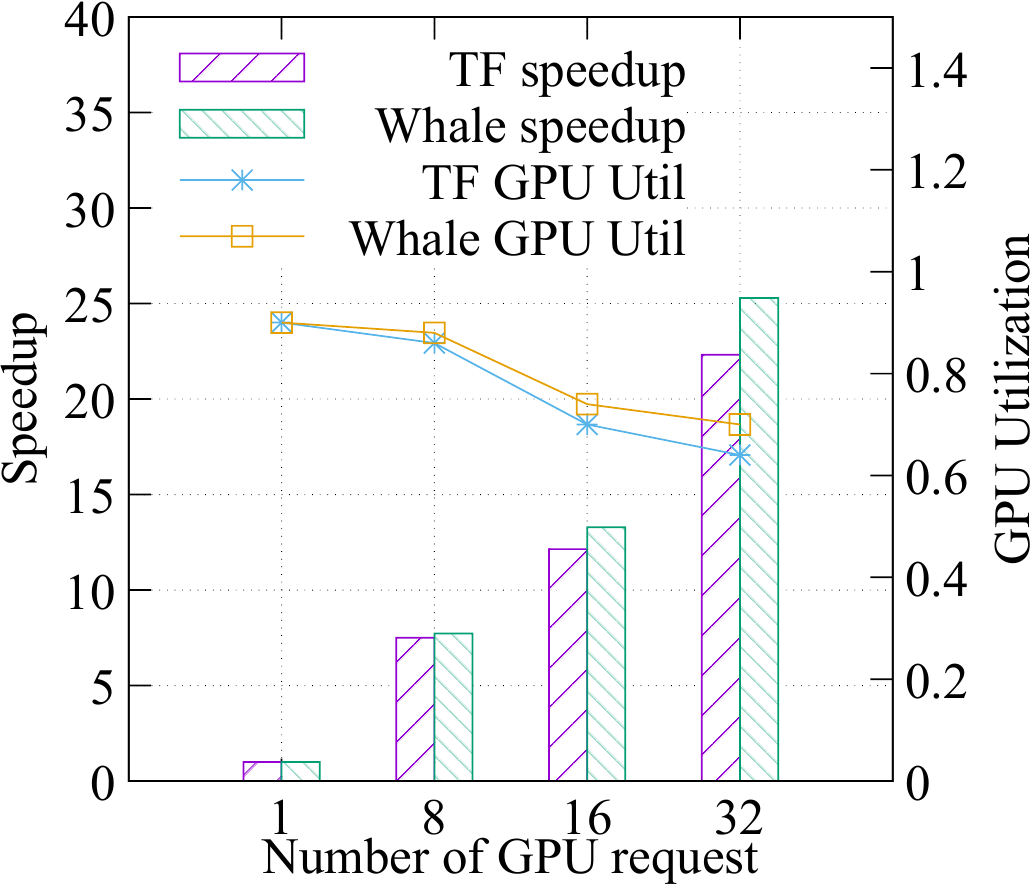}
    \caption{\whale{} DP vs TF DP on ResNet.}
    \label{fig:compare-tf-dp-resnet}
	\end{minipage}
  \hfill
  \begin{minipage}[t]{0.24\textwidth} %
    \centering
    \includegraphics[width=\textwidth]{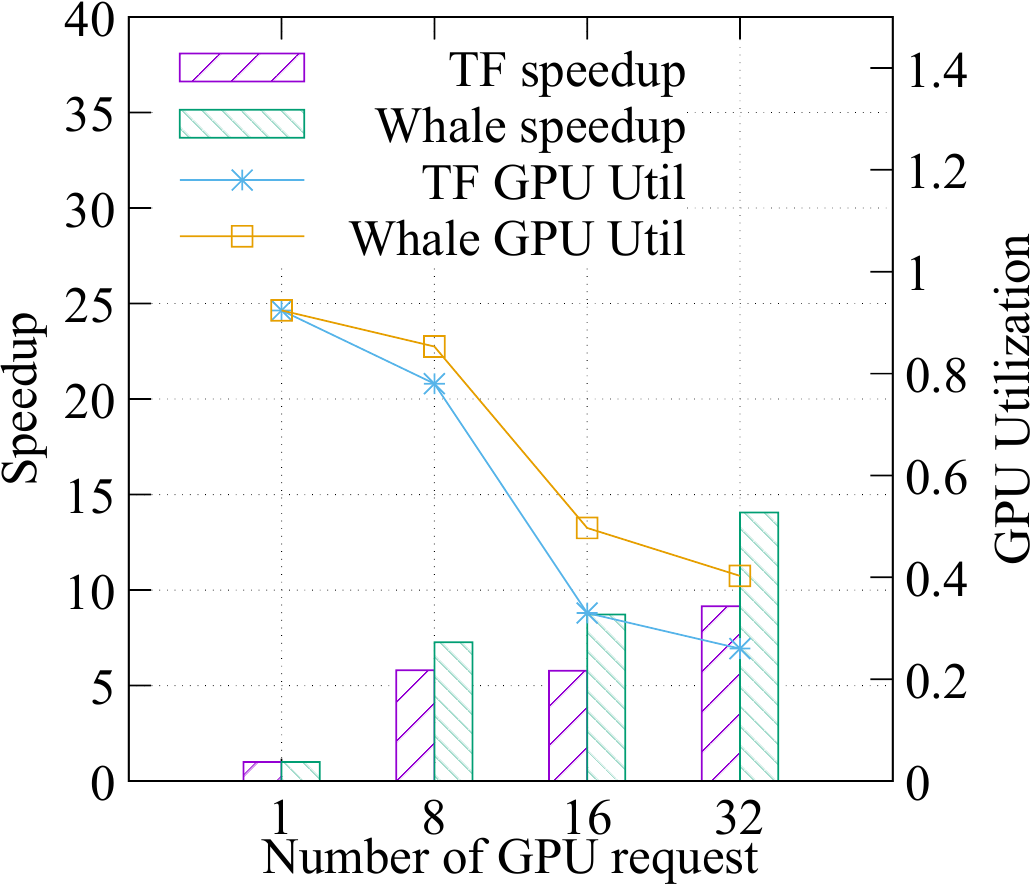}
    \caption{\whale{} DP vs TF DP on BertLarge.}
    \label{fig:compare-tf-dp-bert}
	\end{minipage}
  \hfill
	\begin{minipage}[t]{0.24\textwidth} %
		\centering 
    \includegraphics[width=\linewidth]{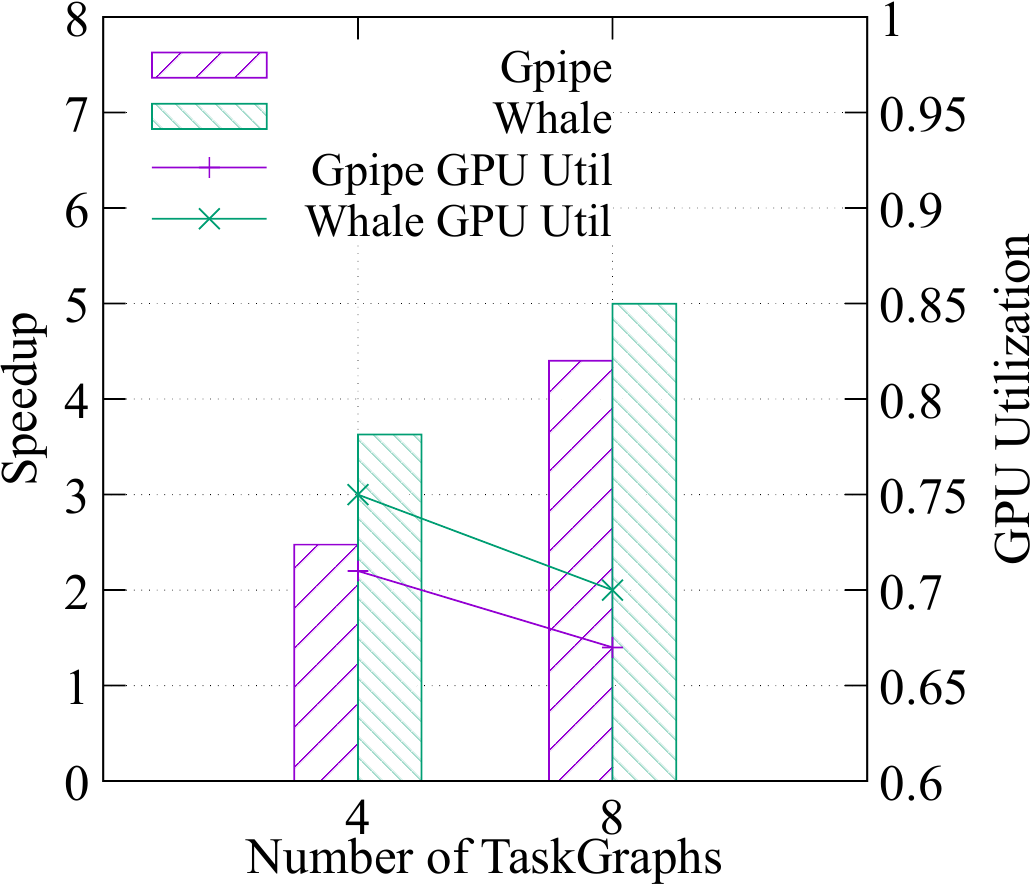}
    \caption{\whale{} Pipeline vs GPipe.}
    \label{fig:compare-gpipe}
	\end{minipage}
  \hfill
  \begin{minipage}[t]{0.24\textwidth} %
    \centering
    \includegraphics[width=\linewidth]{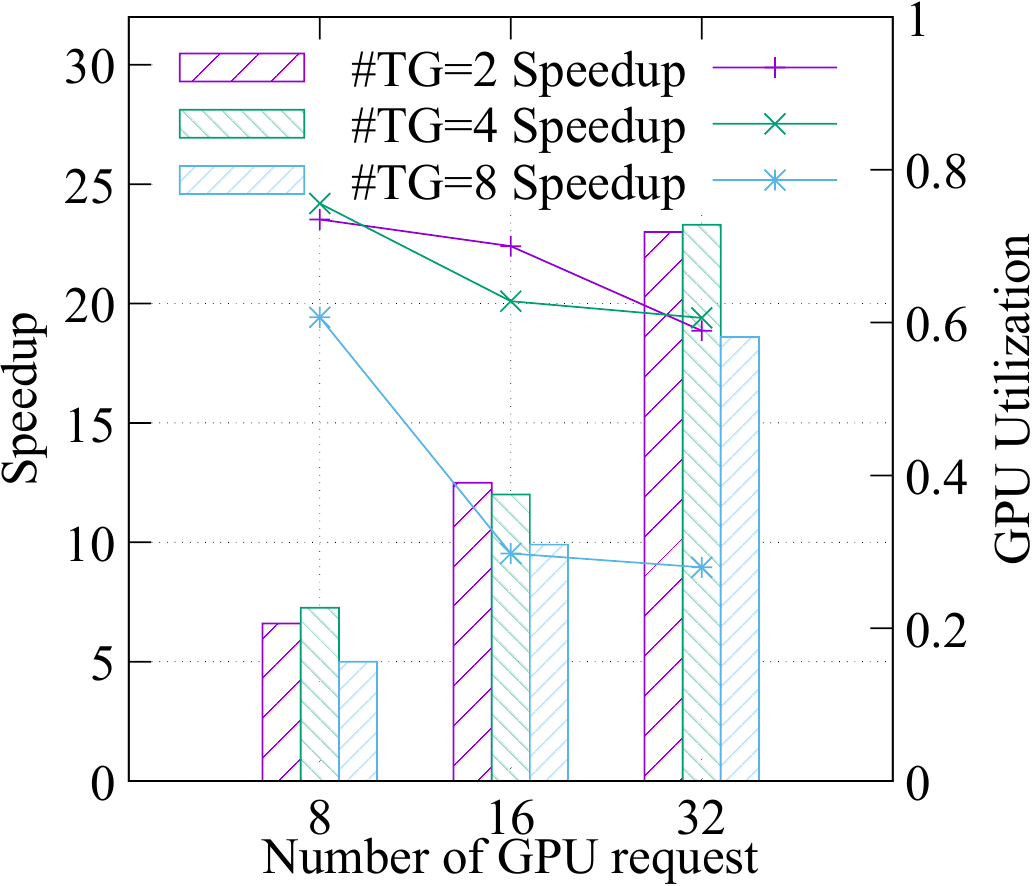}
    \caption{Hybrid pipeline parallelism on BertLarge.}
    \label{fig:epl-pipeline-stage}
	\end{minipage}
	\vspace{-0.15in}
\end{figure*}

\section{Implementation}
\label{sec:impl}
\whale{} is implemented as a standalone library without modification of the deep learning framework,
which is compatible with TensorFlow1.12 and TensorFlow1.15\cite{abadi2016tensorflow}.
The source code of \whale{} includes 13179 lines of Python code and 1037 lines of C++ code.
We have open-sourced\footnote{\url{https://github.com/alibaba/EasyParallelLibrary}} the \whale{} framework to help giant model training accessible to more users.

\whale{} enriches the local model with augmented information such as phase information, parallelism annotation, \etc,
which is crucial to parallelism implementation.
To assist the analysis of the user model without modifying the user code,
\whale{} inspects and overwrites TensorFlow build-in functions to capture augmented information.
For example, operations are marked as backward when $tf.gradients$ or $compute\_gradients$ functions are called.

The parallel strategy is implemented by rewriting the computation graph.
We implement a general graph editor module for ease of graph rewriting,
 which includes functions such as subgraph clone, node replacement, dependency control, and so on.
To implement data parallelism, \whale{} first clones all operations and tensors defined in a local \taskgraph{} and
 replaces the device for model replicas.
  Then it inserts NCCL\cite{nccl2019} AllReduce\cite{sergeev2018horovod} operation to synchronize gradients for each \taskgraph{} replica.
To implement tensor model parallelism, \whale{} shards the \taskgraph{} by matching a series of predefined patterns, replacing them with corresponding distributed implementation,
and inserting communication operations as needed.
To implement pipeline parallelism, \whale{} builds a pipeline strategy module that supports state-of-the-art strategies\cite{huang2018gpipe,narayanan2019pipedream,fan2020dapple}.
By default, \whale{} adopts a backward-first strategy which is similar to PipeDream\cite{narayanan2019pipedream}.
The pipeline strategy is implemented by first partitioning the minibatch into micro-batches.
The interleaving of forward-backward micro-batch execution is achieved by inserting control dependency operations among entrance and exit operations of different \taskgraph{}s.

To assist hardware-aware optimizations, \whale{} implements profiling tools that profile the model FLOPS and peak memory consumption.
The parallel planner gets the hardware information from our internal GPU cluster,
which is used to generate an efficient parallel plan by balancing the computing workloads over heterogeneous GPUs.

Besides, \whale{} is highly optimized in both computing efficiency and memory utilization by integrating with
a series of optimization technologies such as ZERO\cite{rajbhandari2020zero}, recomputation\cite{chen2016training},
 CPU offload\cite{ren2021zero}, automatic mixed precision\cite{micikevicius2017mixed}, communication optimization\cite{sergeev2018horovod},
  XLA\cite{abadi2016tensorflow}, etc.

\section{Experiment}
\label{sec:exp}

In this section, we first demonstrate the efficiency of the parallelism strategy by evaluating micro-benchmarks.
We then evaluate the training with heterogeneous GPUs to show the advantages of the hardware-aware load balance algorithm.
We end by showing the effectiveness and efficiency of \whale{} by two industry-scale multimodal model training cases.
All the experiments are conducted on a shared cloud GPU cluster.
Every cluster node is equipped with a 96-core Intel Xeon Platinum 8163 (Skylake) @2.50GHz with 736GB RAM, running CentOS 7.7.
Each node consists of 2/4/8 GPUs, with NVIDIA 32-GB V100 GPUs\cite{nvidiav100} or NVIDIA 16-GB P100 GPUs\cite{nvidiap100}, powered by NVIDIA driver 418.87, CUDA 10.0, and cuDNN 7. Nodes are connected by 50Gb/s ethernet.
All the models are implemented based on TensorFlow 1.12.





\subsection{Micro-benchmark}

In this section, we evaluate \whale{} with a series of micro-benchmarks.
We first demonstrate that \whale{} is efficient in single parallel strategy by
 comparing with TensorFlow Estimator\cite{dillon2017tensorflow} DP and GPipe\cite{huang2018gpipe} pipeline.
We then show the advantages of \whale{} hybrid strategies over single parallel strategy.
Next, we measure the overhead of the bridge layer for hybrid strategies.
Finally, we evaluate the effect of sharding patterns in automatic \taskgraph{} partitioning.

\subsubsection{Performance of Single Parallel Strategy}

We evaluate \whale{} DP by comparing it with TensorFlow Estimator DP, using the BertLarge\cite{devlin2018bert} and ResNet50\cite{he2016deep}
 on different number of V100 GPUs.
 Figure~\ref{fig:compare-tf-dp-resnet} and Figure~\ref{fig:compare-tf-dp-bert} show the training throughput speedup on ResNet50 and BertLarge respectively.
 The throughput speedup is calculated by dividing the training throughput on $N$ devices by the throughput on one device.
 \whale{} DP consistently obtained better speedup and higher GPU utilization than TensorFlow Estimator DP.
 Such findings could be attributed to \whale{}'s communication optimization technologies such as hierarchical and grouped AllReduce, which is similar to Horovod\cite{sergeev2018horovod}.
 

We then evaluate the efficiency of \whale{} pipeline parallelism  by comparing with GPipe\cite{huang2018gpipe}.
The pipeline scheduling strategy in \whale{} is similar to PipeDream\cite{narayanan2019pipedream}.
The experiments are conducted using the BertLarge model with 4/8 pipeline stages on the different numbers of V100 GPUs.
As shown in Figure~\ref{fig:compare-gpipe}, the training throughput speedup of \whale{} outperforms GPipe in both 4 stages and 8 stages by 1.45X and 1.14X respectively.
We attribute the performance gain to the use of the alternating forward-backward scheduling policy\cite{narayanan2019pipedream}, which improves GPU utilization.
We also find that the pipeline performance is sensitive to the $num\_task\_graph$,
 thus exposing it as a configurable parameter can help achieve a better performance when models and computing resources change.

\begin{figure*}[t]
	\begin{minipage}[t]{0.24\textwidth}
		\centering 
		\includegraphics[width=0.95\linewidth]{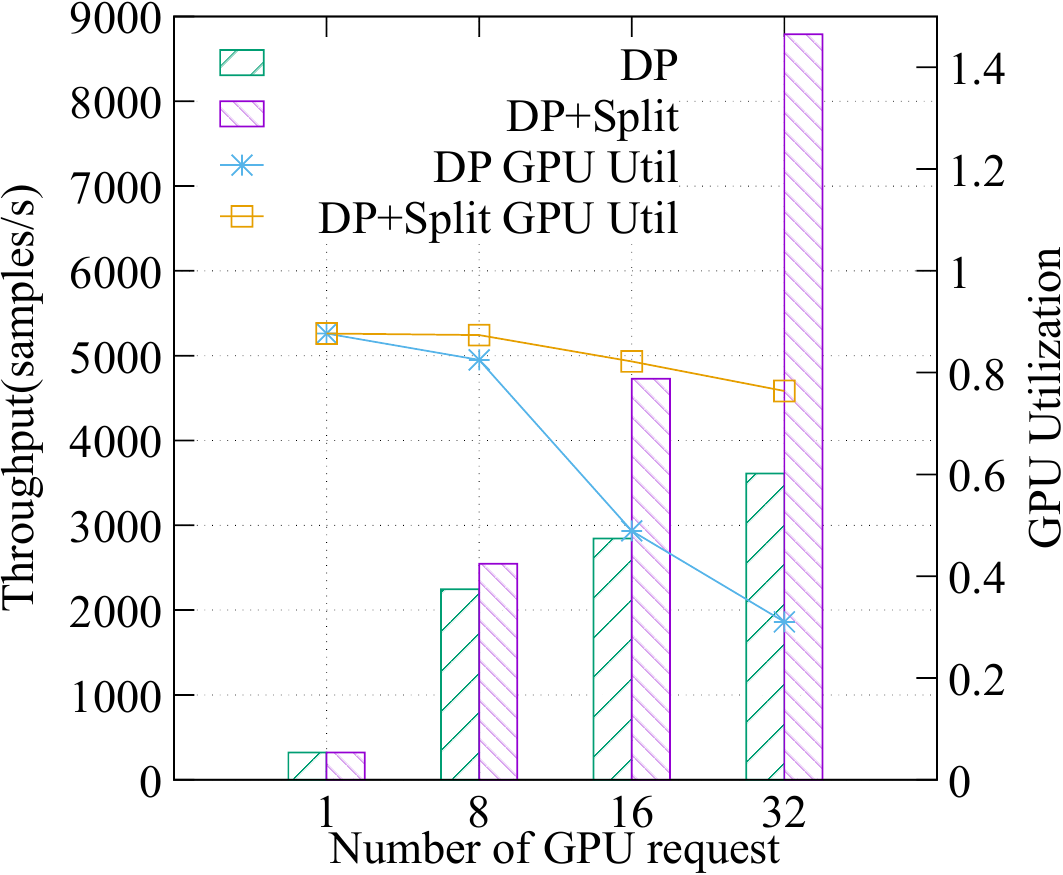} 
		\vspace{-0.1in}
		\caption{DP vs Hybrid on ResNet50 w/ 100K classes.}
		\label{fig:resnet-10w} 
	\end{minipage}
  \hfill
	\begin{minipage}[t]{0.24\textwidth}
		\centering 
		\includegraphics[width=0.95\linewidth]{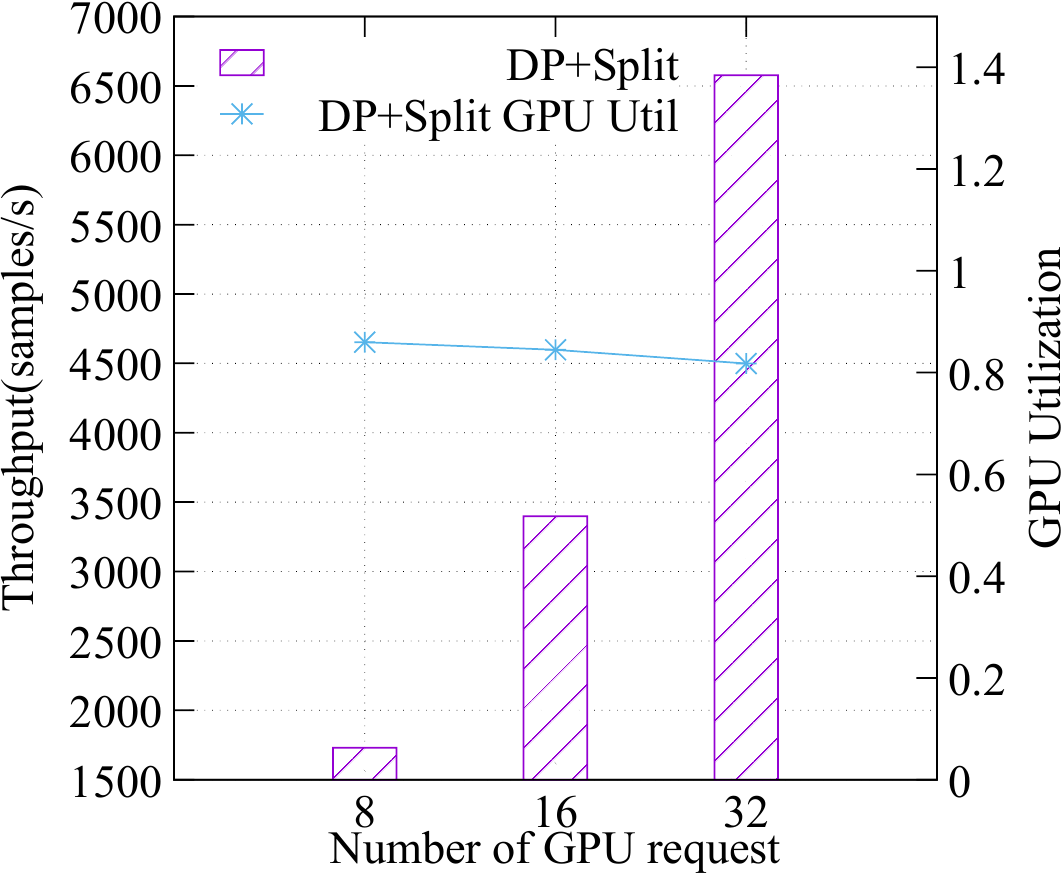} 
		\vspace{-0.1in}
		\caption{Hybrid strategy on ResNet50 w/ 1M classes.} 
		\label{fig:resnet-100w} 
	\end{minipage}%
  \hfill
	\begin{minipage}[t]{0.24\textwidth} 
		\centering 
		\includegraphics[width=0.95\linewidth]{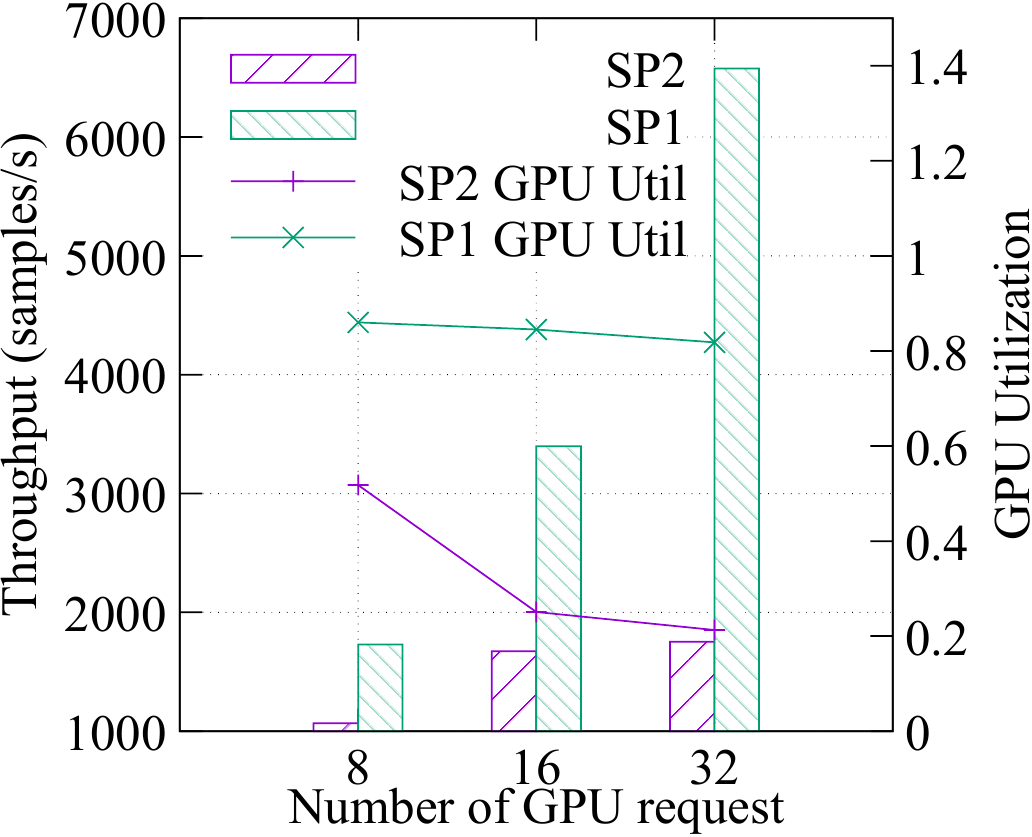} 
		\vspace{-0.1in}
		\caption{Effect of Sharding Pattern.} 
		\label{fig:exp-sharding-pattern} 
	\end{minipage} 
  \hfill
  \begin{minipage}[t]{0.24\textwidth} 
		\centering 
		\includegraphics[width=0.8\linewidth]{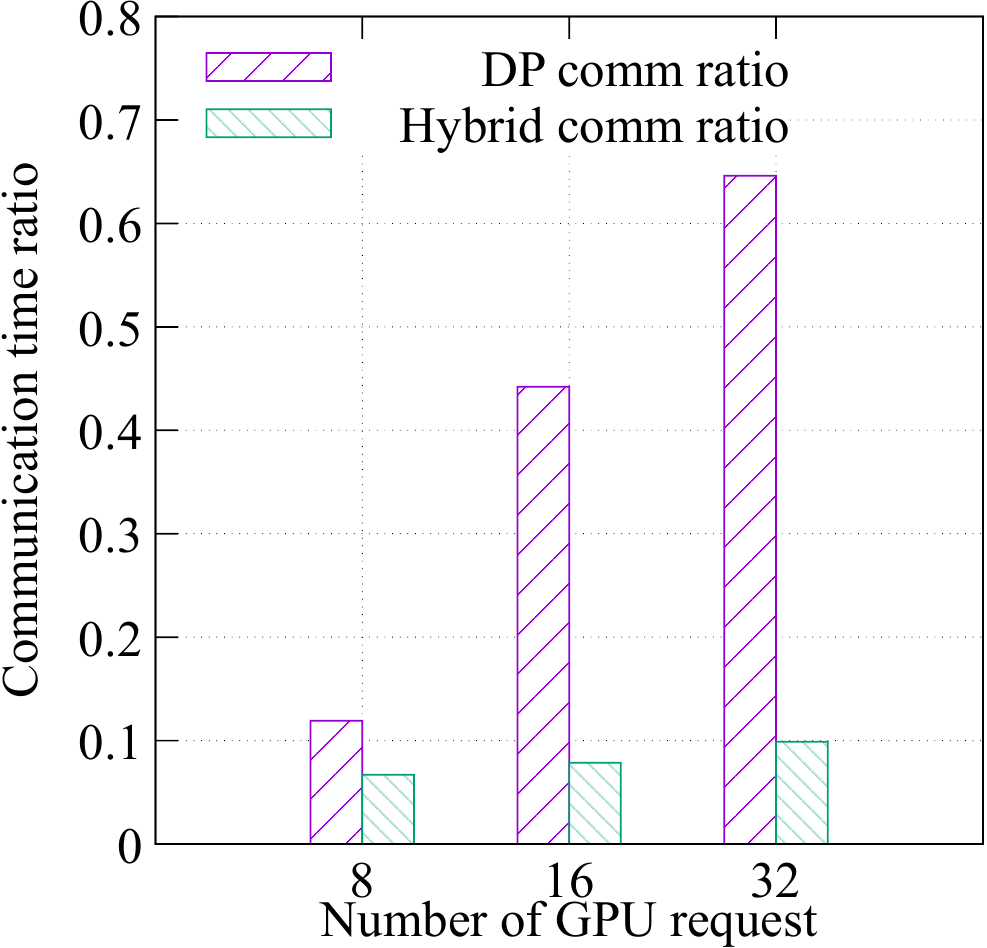}
		\vspace{-0.1in}
		\caption{Overhead of Bridge Layer.} 
		\label{fig:bridge-overhead} 
	\end{minipage}
	\vspace{-0.2in}
\end{figure*}


\subsubsection{Performance of Hybrid Strategy}
We evaluate hybrid strategies by comparing them with the single parallel strategy.
We also compare the performances of hybrid strategies on different numbers of devices.
We select two typical types of hybrid strategies:
1) Nested pipeline with DP;
and 2) Combination of DP and  tensor model parallelism.

We first apply a nested pipeline with DP to the BertLarge model on V100 GPUs.
The model is partitioned into 2/4/8 number of \taskgraph{}s, and we measure the training performance of each model on 8/16/32 GPUs.
Figure~\ref{fig:epl-pipeline-stage} shows that pipelines with 2 \taskgraph{}s and 4 \taskgraph{}s get similar training speedups and GPU utilization.
However, we observe a performance drop on 8 \taskgraph{}s and lower GPU utilization compared to 2/4 \taskgraph{}s.
This is because 8 \taskgraph{}s lead to relatively fewer model operations in each \taskgraph{},
and the GPU computation is not enough to overlap the inter-\taskgraph{} communication, resulting in poor performance.


Next, we evaluate the combination hybrid strategy on a large-scale image classification model,
 as we have discussed in Section~\ref{sec:strategy} and illustrated in Figure~\ref{fig:hybrid-parallel}.
 We perform experiments on classification numbers 100K and 1M on different numbers of V100 GPUs.
 To reduce the communication overhead of hybrid parallelism, we collocate the $ResNet50$ replicas with $FC$ partitions.
 We compare the hybrid results of 100K classes with DP, as shown in Figure~\ref{fig:resnet-10w}, hybrid parallelism outperforms data parallelism by 1.13X, 1.66X, and 2.43X training throughput speedup with 8, 16, and 32 GPUs respectively,
with the line plot corresponding to GPU utilization.
When the number of workers increases, hybrid parallelism maintains a near-linear speedup, while the DP strategy fails drastically beyond 16 workers. 
This is because the heavy $FC$ layer (the parameter size of ResNet50
backbone is 90 MB, while the parameter size of $FC$ layer is 782MB) incurs a huge gradient synchronization overhead.
For the task of 1M classes, DP fails due to OOM. 
 With hybrid parallelism, \whale{} allows for  the training of image classification task with one million classes.
Figure~\ref{fig:resnet-100w} shows the performance of hybrid parallelism over 8/16/32 GPUs.
The training throughputs from 8 GPUs to 32 GPUs achieve 95\% scaling efficiency, which highlights the need for using a hybrid strategy.

\subsubsection{Overhead of Bridge Layer}
To demonstrate the efficiency of the hybrid strategy,
We measure the overhead of the bridge layer by profiling the bridge layer time with 100K classes on 8/16/32 GPUs.
We then compare the overhead of gradient AllReduce time in DP with the bridge overhead to understand the performance gain from hybrids.
As shown in Figure~\ref{fig:bridge-overhead}, the overhead of the bridge layer takes around 6\% in overall training time in 8 GPUs and 10\% in 32 GPUs.
The overhead of the hybrid is reduced by 6X on 32 GPUs compared to gradient synchronization overhead of pure DP.

\subsubsection{Effect of Sharding Pattern}

As \whale{} automatically chooses a sharding pattern with minimum communication cost (Section~\ref{sec:partition}),
to demonstrate the effect of exploring the sharding patterns, we force the framework to use a specific pattern in this experiment.
We evaluate two types of sharding patterns as illustrated in Figure~\ref{fig:sharding-pattern} on large scale image task with 100K classes.
$SP1$ shards the second input tensor in the second tensor dimension, and $SP2$ shards the two input tensors and aggregates the results with $AllReduce$.
The comparison results of the two sharding patterns are shown in Figure~\ref{fig:exp-sharding-pattern},
 where $SP1$ outperforms $SP2$ by 1.6X to 3.75X as the number of requested GPUs increases from 8 to 32, as $SP1$ has a lower communication cost than $SP2$.
 The exploration of sharding patterns allows for the possibility of system optimization in distributed model implementation.

\subsection{Performance of Load Balance}
We show the benefits of the hardware-aware load balancing algorithm by evaluating \datap{} and \pipe{}.

\begin{figure}[t]
	\begin{minipage}[t]{0.47\linewidth} %
		\centering 
		\includegraphics[width=\textwidth]{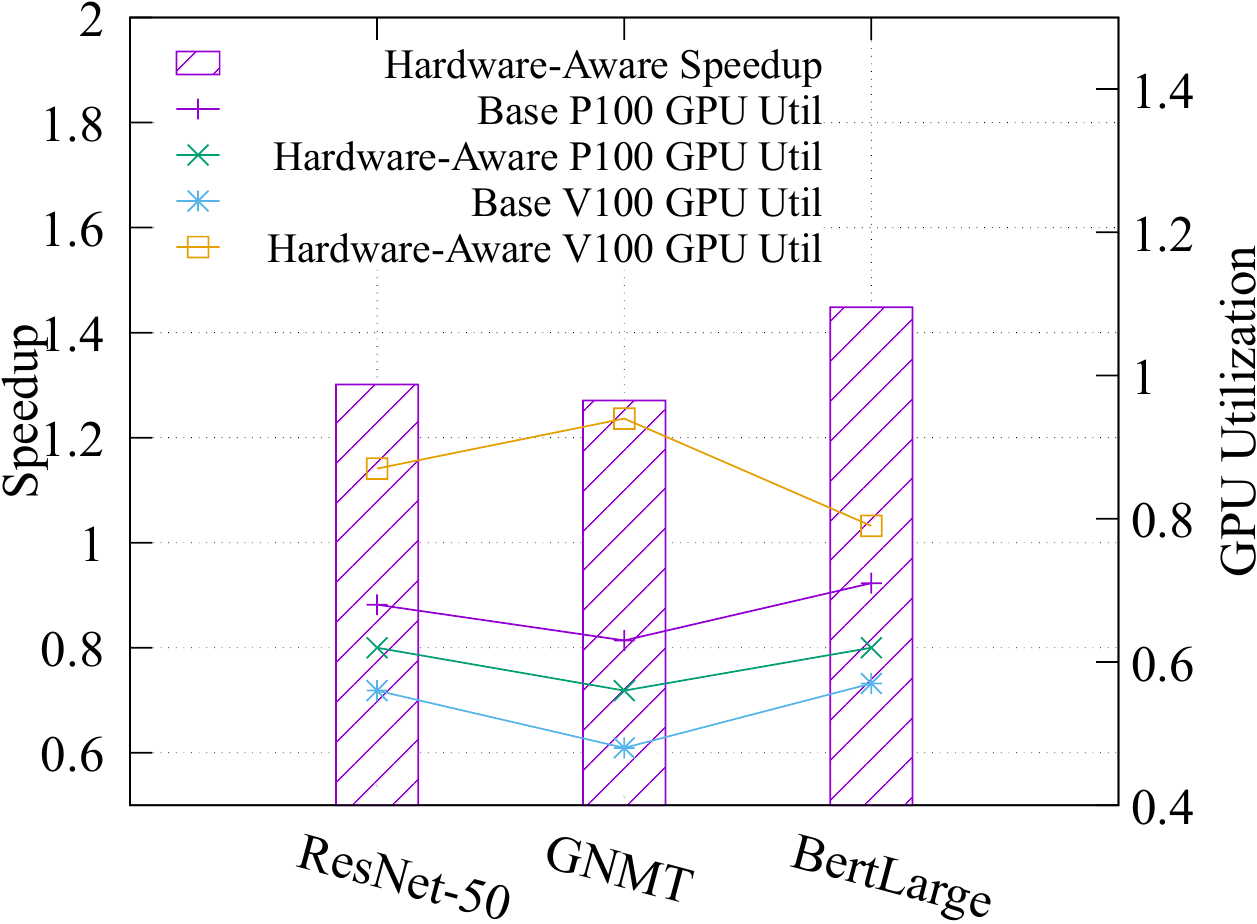}
		\vspace{-0.25in}
    \caption{Hardware-Aware Data Parallelism.}
    \label{fig:heto-dp}
	\end{minipage}
	\begin{minipage}[t]{0.47\linewidth} %
		\centering 
		\includegraphics[width=\textwidth]{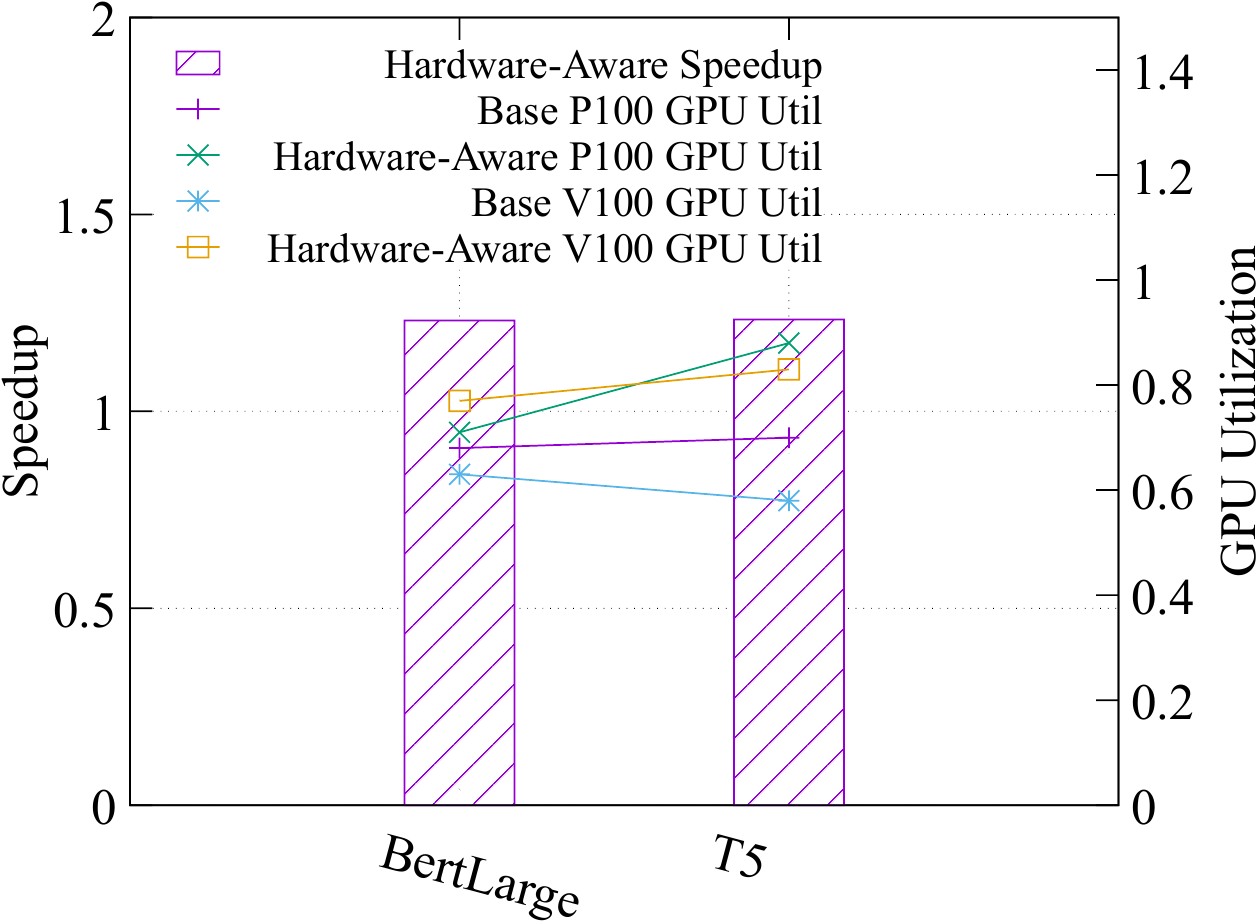}
		\vspace{-0.25in}
    \caption{Hardware-Aware Pipeline Parallelism.}
    \label{fig:heto-pipe}
	\end{minipage}
	\vspace{-0.15in}
\end{figure}

For \datap{}, we evaluate three typical models, including ResNet50, BertLarge, and GNMT\cite{wu2016google}.
 The experiments are conducted on heterogeneous GPUs that consist of 8 32GB V100 GPUs and 8 16GB P100 GPUs.
We set the same batch size for all model replicas as the baseline. 
 We then apply the hardware-aware algorithm to each model and get the speedup compared with the baseline performance, as shown in  Figure~\ref{fig:heto-dp}.
 \whale{} outperforms the baseline in all three models by a factor from 1.3X to 1.4X. We also measure GPU utilization
 and report the average metric for each GPU type. The hardware-aware policy significantly improves the GPU utilization of V100 by 1.39X to 1.96X for the three models,
 which improves the overall training performance.

  For \pipe{}, we evaluate two models, including BertLarge and T5-Large\cite{xue2020mt5}.
   The training is performed on heterogeneous GPUs that consist of 4 32GB V100 GPUs and 4 16GB P100 GPUs.
   Both BertLarge and T5-Large are partitioned into 4 stages. We further apply a nested DP to $pipeline$.
   We set the evenly partitioned model as the baseline.
   We conducted training with the hardware-aware policy and got about 20\% speedup on both models, as shown in  Figure~\ref{fig:heto-pipe}. 
   The GPU utilization of hardware-aware load balancing strategy improved the GPU utilization of V100 by around 40\%,
   which shows the efficiency of the hardware-aware load balancing algorithm.   
\subsection{Industry-Scale Giant Model Training}
\subsubsection{Training M6-10B Model}



\begin{figure}[t]
	\centering
\begin{minipage}[t]{0.5\linewidth}
	\vspace{-5.5\baselineskip}
\centering
\begin{lstlisting}[style = Python, caption=M6-10B model with pipeline, belowcaptionskip = 0pt, captionpos=b, label=case:m610b]
import whale as wh
wh.init(wh.Config({
  "num_micro_batch": 35,
  "num_task_graph": 8}))
# Define M6 model.
m6_model_def()
\end{lstlisting}
\end{minipage}
		  \hfill
	\begin{minipage}[t]{0.38\linewidth} %
		\centering 
		\includegraphics[width=\linewidth]{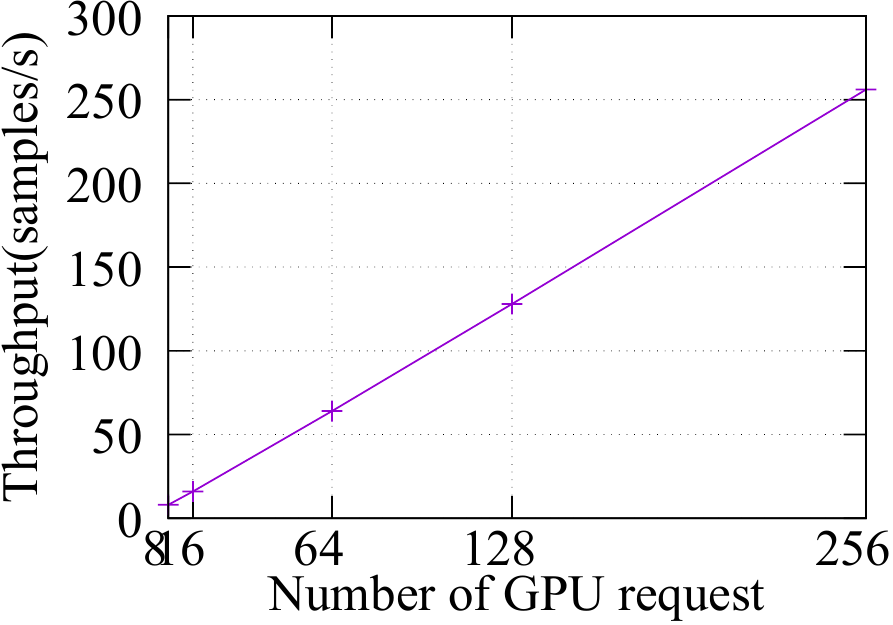} 
		\vspace{-0.3in} 
		\caption{M6-10B with Pipeline and DP.}
		\label{fig:pipeline-exp}
	\end{minipage}
	\vspace{-0.2in}
\end{figure}

 The M6-10B\cite{lin2021m6} model is a Chinese multimodal model with 10 billion parameters.
 The model consists of 24 encoder layers and 24 decoder layers. We use Adafactor\cite{shazeer2018adafactor} as the training optimizer.
 We parallelize the training of M6-10B model with a hybrid parallel strategy, by nesting \pipe{} and \datap{}.
 \whale{} can easily scale a local M6 model to a distributed one by only adding a few lines on top of the model definition as shown in Example~\ref{case:m610b}.
We set the number of pipeline stages to 8 and the number of micro-batches to 35.
We enable recomputation\cite{chen2016training} to save activation memory during training.
The training performance is evaluated on 32-GB V100 GPUs. Each node contains 8 GPUs.
When scaling the computing nodes from 8 to 32, \whale{} achieved 91\% scalability, as shown in Figure~\ref{fig:pipeline-exp}.

\subsubsection{Training M6-MoE Model to Trillions}
\label{sec:exp-m6}
We scale the model parameters to 10 trillion (10T) by switching to hybrids of $DP$ and tensor model parallelism with only a small number of lines of code change.
 The computation cost of training dense models is proportional to the model parameters.
  If we scale the dense 10B model to the dense 10T model linearly without considering overhead,
   we need at least 256,000 NVIDIA V100 GPUs.
 Instead of scaling the M6 model with dense structure, we adopt M6-MoE\cite{yang2021exploring} model with sparse expert solution\cite{lepikhin2020gshard,fedus2021switch}.
The sample code of the MoE structure is implemented with \whale{} by adding four lines, as shown in Example~\ref{case:moe}.
Line~3 sets the default parallel primitive as $replicate$, i.e., \datap{} is applied for the operations if not explicitly annotated.
Line~5 partitions the computation defined under $split$ scope across devices.
 

\begin{minipage}{0.9\linewidth}
\centering
\begin{lstlisting}[style = Python, caption=Distributed MoE model, label=case:moe, belowcaptionskip = 0pt, captionpos=b, numbers=left]
import whale as wh
wh.init()
wh.set_default_strategy(wh.replicate(total_gpus))
combined_weights,dispatch_inputs=gating_dispatch()
with wh.split(total_gpus):
  outputs = MoE(combined_weights, dispatch_inputs)
\end{lstlisting}
\end{minipage}

We evaluate M6-MoE model with 100 billion, 1 trillion and 10 trillion parameters respectively, the detailed configurations can be found in \cite{yang2021exploring,lin2021m610t}.
We enable built-in technologies of \whale{} to optimize the training process,
 such as recomputation\cite{chen2016training}, AMP (auto mixed precision)\cite{nvidiaamp}, XLA\cite{tfxla}, CPU offloading\cite{ren2021zero}, etc.
We can train the M6-MoE-100B model with 100 million samples on 128 V100 in 1.5 days.
We advance the model scale to 1 trillion parameters on solely 480 NVIDIA V100 GPUs, in comparison with the recent SOTA on 2048 TPU cores\cite{fedus2021switch}.
 We further scale the model to 10 trillion parameters by adopting optimized tensor offloading strategies\cite{lin2021m610t} with 512 NVIDIA V100 GPUs.
\whale{} can scale models from 100 billion to 10 trillion without code changes,
which makes giant model training accessible to most users.

\section{Related Work}

\paragraph{Giant model training.}
TensorFlow\cite{abadi2016tensorflow} and PyTorch\cite{paszke2019pytorch} provide well-supported data parallelism and vanilla model parallelism by explicitly assigning operations to specific devices.
However, they are not efficient enough for giant model training.
Megatron~\cite{shoeybi2019megatron}, GPipe~\cite{huang2018gpipe}, and Dapple~\cite{fan2020dapple} 
have proposed new parallel training strategies to scale the training of large-scale models.
DeepSpeed~\cite{rasley2020deepspeed} lacks general support for tensor model parallelism, besides, model layers are required to rewrite in sequential for pipeline parallelism. 
GShard~\cite{lepikhin2020gshard} supports operator splitting by introducing model weight annotations and tensor dimension specifications.
The high performance of those works is achieved by exposing low-level system abstractions to users (\eg, device placement, equivalent distributed implementation for operators),
or enforcing model or tensor partition manually, which results in significant user efforts.
As a parallel work to \whale{}, GSPMD~\cite{xu2021gspmd} extends GShard by annotating tensor dimensions mapping 
for both automatic and manual operator partitioning.
As a general giant model training framework, \whale{} adopts a unified abstraction to express different parallel strategies and their hybrid nests and combinations, 
utilizing high-level annotations and pattern matching for operator splitting.
\whale{} further scales to M6-10T through automatically distributed graph optimizations with the awareness of heterogeneous resources.

Zero\cite{rajbhandari2020zero, ren2021zero, rajbhandari2021zero}
optimizes memory usage by removing redundant GPU memory, offloading computation to the CPU host, and utilizing non-volatile memory respectively.
Recomputation\cite{chen2016training} trades computation for memory by recomputing tensors from checkpoints. 
Such memory optimization approaches are orthogonal to \whale{}, which can be further combined for giant model training efficiently.

\vspace{-0.2in}
\paragraph{Graph optimization.}
Deep learning is powered by dataflow graphs with optimizations to rewrite the graph for better performance,
such as TensorFlow XLA~\cite{abadi2016tensorflow}, TVM~\cite{tvm2020}, Ansor~\cite{zheng2020ansor}, AStitish~\cite{zheng2022astitch}, \etc{}
TASO~\cite{jia2019taso} and PET~\cite{wang2021pet} adopt a graph substitution approach to optimize the computation graph automatically. 
Those works mainly focus on the performance of a single GPU, 
while \whale{} utilizes the graph optimization approach for achieving efficient performance in distributed training.
Tofu~\cite{wang2019tofu} and SOAP~\cite{jia2019soap} also use graph partition to 
produce distributed execution plans, but with a high search cost.
\whale{} utilizes the introduced annotations to shrink the search space, thus making graph optimization practical for giant model training at a trillion scale. 
Besides, \whale{} extends the graph optimization approach to complicated parallel strategies in a unified abstraction,
 capable of pipeline parallelism, tensor model parallelism, and hybrid parallelism.

\vspace{-0.2in}
\paragraph{Resource heterogeneity.}
Philly~\cite{jeon2019analysis} reports the trace study in multi-tenant GPU clusters of Microsoft and 
shows the effect of gang scheduling on job queuing.
MLaaS~\cite{weng2022mlasas} studies a two-month trace of a heterogeneous GPU cluster in Alibaba PAI. 
Gandiva~\cite{gandiva2018} shows jobs are different in sensitivity to allocated resources.
\whale{} is capable of adapting to resource heterogeneity, which can reduce the queuing delay of 
giant model training with hundreds of GPUs. 
The design of \whale{} advocates the approach of decoupling model programming and distributed execution.
It dynamically generates an efficient execution plan by considering the properties of both model and heterogeneous resources.

\section{Conclusion}

\whale{} demonstrates the possibility of achieving efficiency, programmability, and adaptability 
in a scalable deep learning framework for training trillion-parameter models.
\whale{} supports various parallel strategies using a unified abstraction, 
hides distributed execution details through new primitive annotations,
and adapts to heterogeneous GPUs with automatic graph optimizations.
Going forward, we hope that \whale{} can become a large-scale deep learning training foundation
to further engage model algorithm innovations and system optimizations in parallel, 
making giant model training technology to be adopted easily and efficiently at scale.

\section*{Acknowledgements}
We would like to thank our anonymous shepherd and reviewers for their valuable comments and
suggestions. We would also like to thank the M6 team and all users of \whale{} for their help and suggestions.

\bibliographystyle{plain}
\bibliography{reference}

\begin{thebibliography}{10}

\bibitem{nvidiaamp}
Automatic mixed precision for deep learning.
\newblock \url{https://developer.nvidia.com/automatic-mixed-precision}.

\bibitem{nvidiap100}
Nvidia tesla p100.
\newblock \url{https://www.nvidia.com/en-us/data-center/tesla-p100/}.

\bibitem{nvidiav100}
Nvidia v100 tensor core gpu.
\newblock \url{https://www.nvidia.com/en-us/data-center/v100/}.

\bibitem{nvlink}
{NVLink}.
\newblock \url{https://www.nvidia.com/en-us/data-center/nvlink/}.

\bibitem{tfxla}
Xla: Optimizing compiler for machine learning.
\newblock \url{https://www.tensorflow.org/xla}.

\bibitem{nccl2019}
Nccl.
\newblock \url{https://developer.nvidia.com/nccl}, 2019.

\bibitem{abadi2016tensorflow}
Mart{\'\i}n Abadi, Paul Barham, Jianmin Chen, Zhifeng Chen, Andy Davis, Jeffrey
  Dean, Matthieu Devin, Sanjay Ghemawat, Geoffrey Irving, Michael Isard, et~al.
\newblock Tensorflow: A system for large-scale machine learning.
\newblock In {\em 12th {USENIX} symposium on operating systems design and
  implementation ({OSDI} 16)}, pages 265--283, 2016.

\bibitem{brown2020language}
Tom~B Brown, Benjamin Mann, Nick Ryder, Melanie Subbiah, Jared Kaplan, Prafulla
  Dhariwal, Arvind Neelakantan, Pranav Shyam, Girish Sastry, Amanda Askell,
  et~al.
\newblock Language models are few-shot learners.
\newblock {\em arXiv preprint arXiv:2005.14165}, 2020.

\bibitem{tvm2020}
Tianqi Chen, Thierry Moreau, Ziheng Jiang, Lianmin Zheng, Eddie Yan, Haichen
  Shen, Meghan Cowan, Leyuan Wang, Yuwei Hu, Luis Ceze, Carlos Guestrin, and
  Arvind Krishnamurthy.
\newblock {TVM}: An automated end-to-end optimizing compiler for deep learning.
\newblock In {\em 13th {USENIX} Symposium on Operating Systems Design and
  Implementation ({OSDI} 18)}, pages 578--594, Carlsbad, CA, October 2018.
  {USENIX} Association.

\bibitem{chen2016training}
Tianqi Chen, Bing Xu, Chiyuan Zhang, and Carlos Guestrin.
\newblock Training deep nets with sublinear memory cost.
\newblock {\em arXiv preprint arXiv:1604.06174}, 2016.

\bibitem{adam2014}
Trishul Chilimbi, Yutaka Suzue, Johnson Apacible, and Karthik Kalyanaraman.
\newblock Project adam: Building an efficient and scalable deep learning
  training system.
\newblock In {\em 11th USENIX Symposium on Operating Systems Design and
  Implementation (OSDI 14)}, pages 571--582, Broomfield, CO, October 2014.
  USENIX Association.

\bibitem{distbelief2012}
Jeffrey Dean, Greg~S. Corrado, Rajat Monga, Kai Chen, Matthieu Devin, Quoc~V.
  Le, Mark~Z. Mao, Marc’Aurelio Ranzato, Andrew Senior, Paul Tucker, Ke~Yang,
  and Andrew~Y. Ng.
\newblock Large scale distributed deep networks.
\newblock In {\em NIPS}, 2012.

\bibitem{devlin2018bert}
Jacob Devlin, Ming-Wei Chang, Kenton Lee, and Kristina Toutanova.
\newblock Bert: Pre-training of deep bidirectional transformers for language
  understanding.
\newblock {\em arXiv preprint arXiv:1810.04805}, 2018.

\bibitem{dillon2017tensorflow}
Joshua~V Dillon, Ian Langmore, Dustin Tran, Eugene Brevdo, Srinivas Vasudevan,
  Dave Moore, Brian Patton, Alex Alemi, Matt Hoffman, and Rif~A Saurous.
\newblock Tensorflow distributions.
\newblock {\em arXiv preprint arXiv:1711.10604}, 2017.

\bibitem{dosovitskiy2020image}
Alexey Dosovitskiy, Lucas Beyer, Alexander Kolesnikov, Dirk Weissenborn,
  Xiaohua Zhai, Thomas Unterthiner, Mostafa Dehghani, Matthias Minderer, Georg
  Heigold, Sylvain Gelly, et~al.
\newblock An image is worth 16x16 words: Transformers for image recognition at
  scale.
\newblock {\em arXiv preprint arXiv:2010.11929}, 2020.

\bibitem{fan2020dapple}
Shiqing Fan, Yi~Rong, Chen Meng, Zongyan Cao, Siyu Wang, Zhen Zheng, Chuan Wu,
  Guoping Long, Jun Yang, Lixue Xia, Lansong Diao, Xiaoyong Liu, and Wei Lin.
\newblock Dapple: A pipelined data parallel approach for training large models,
  2020.

\bibitem{fedus2021switch}
William Fedus, Barret Zoph, and Noam Shazeer.
\newblock Switch transformers: Scaling to trillion parameter models with simple
  and efficient sparsity, 2021.

\bibitem{gao2020estimating}
Yanjie Gao, Yu~Liu, Hongyu Zhang, Zhengxian Li, Yonghao Zhu, Haoxiang Lin, and
  Mao Yang.
\newblock Estimating gpu memory consumption of deep learning models.
\newblock In {\em Proceedings of the 28th ACM Joint Meeting on European
  Software Engineering Conference and Symposium on the Foundations of Software
  Engineering}, pages 1342--1352, 2020.

\bibitem{he2016deep}
Kaiming He, Xiangyu Zhang, Shaoqing Ren, and Jian Sun.
\newblock Deep residual learning for image recognition.
\newblock In {\em Proceedings of the IEEE conference on computer vision and
  pattern recognition}, pages 770--778, 2016.

\bibitem{huang2018gpipe}
Yanping Huang, Youlong Cheng, Ankur Bapna, Orhan Firat, Mia~Xu Chen, Dehao
  Chen, HyoukJoong Lee, Jiquan Ngiam, Quoc~V Le, Yonghui Wu, et~al.
\newblock Gpipe: Efficient training of giant neural networks using pipeline
  parallelism.
\newblock {\em arXiv preprint arXiv:1811.06965}, 2018.

\bibitem{jeon2019analysis}
Myeongjae Jeon, Shivaram Venkataraman, Amar Phanishayee, Junjie Qian, Wencong
  Xiao, and Fan Yang.
\newblock Analysis of large-scale multi-tenant {GPU} clusters for {DNN}
  training workloads.
\newblock In {\em 2019 {USENIX} Annual Technical Conference ({USENIX} {ATC}
  19)}, pages 947--960, 2019.

\bibitem{jia2019taso}
Zhihao Jia, Oded Padon, James Thomas, Todd Warszawski, Matei Zaharia, and Alex
  Aiken.
\newblock Taso: optimizing deep learning computation with automatic generation
  of graph substitutions.
\newblock In {\em Proceedings of the 27th ACM Symposium on Operating Systems
  Principles}, pages 47--62, 2019.

\bibitem{jia2019soap}
Zhihao Jia, Matei Zaharia, and Alex Aiken.
\newblock Beyond data and model parallelism for deep neural networks.
\newblock In Ameet Talwalkar, Virginia Smith, and Matei Zaharia, editors, {\em
  Proceedings of Machine Learning and Systems 2019, MLSys 2019, Stanford, CA,
  USA, March 31 - April 2, 2019}. mlsys.org, 2019.

\bibitem{kaplan2020scaling}
Jared Kaplan, Sam McCandlish, Tom Henighan, Tom~B Brown, Benjamin Chess, Rewon
  Child, Scott Gray, Alec Radford, Jeffrey Wu, and Dario Amodei.
\newblock Scaling laws for neural language models.
\newblock {\em arXiv preprint arXiv:2001.08361}, 2020.

\bibitem{krizhevsky2014one}
Alex Krizhevsky.
\newblock One weird trick for parallelizing convolutional neural networks.
\newblock {\em arXiv preprint arXiv:1404.5997}, 2014.

\bibitem{lepikhin2020gshard}
Dmitry Lepikhin, HyoukJoong Lee, Yuanzhong Xu, Dehao Chen, Orhan Firat, Yanping
  Huang, Maxim Krikun, Noam Shazeer, and Zhifeng Chen.
\newblock Gshard: Scaling giant models with conditional computation and
  automatic sharding.
\newblock {\em arXiv preprint arXiv:2006.16668}, 2020.

\bibitem{pytorchddp2020}
Shen Li, Yanli Zhao, Rohan Varma, Omkar Salpekar, Pieter Noordhuis, Teng Li,
  Adam Paszke, Jeff Smith, Brian Vaughan, Pritam Damania, and Soumith Chintala.
\newblock Pytorch distributed: Experiences on accelerating data parallel
  training.
\newblock {\em Proc. {VLDB} Endow.}, 13(12):3005--3018, 2020.

\bibitem{lin2021m6}
Junyang Lin, Rui Men, An~Yang, Chang Zhou, Ming Ding, Yichang Zhang, Peng Wang,
  Ang Wang, Le~Jiang, Xianyan Jia, Jie Zhang, Jianwei Zhang, Xu~Zou, Zhikang
  Li, Xiaodong Deng, Jie Liu, Jinbao Xue, Huiling Zhou, Jianxin Ma, Jin Yu,
  Yong Li, Wei Lin, Jingren Zhou, Jie Tang, and Hongxia Yang.
\newblock M6: A chinese multimodal pretrainer, 2021.

\bibitem{lin2021m610t}
Junyang Lin, An~Yang, Jinze Bai, Chang Zhou, Le~Jiang, Xianyan Jia, Ang Wang,
  Jie Zhang, Yong Li, Wei Lin, et~al.
\newblock M6-10t: A sharing-delinking paradigm for efficient multi-trillion
  parameter pretraining.
\newblock {\em arXiv preprint arXiv:2110.03888}, 2021.

\bibitem{liu2021swin}
Ze~Liu, Yutong Lin, Yue Cao, Han Hu, Yixuan Wei, Zheng Zhang, Stephen Lin, and
  Baining Guo.
\newblock Swin transformer: Hierarchical vision transformer using shifted
  windows.
\newblock {\em arXiv preprint arXiv:2103.14030}, 2021.

\bibitem{micikevicius2017mixed}
Paulius Micikevicius, Sharan Narang, Jonah Alben, Gregory Diamos, Erich Elsen,
  David Garcia, Boris Ginsburg, Michael Houston, Oleksii Kuchaiev, Ganesh
  Venkatesh, et~al.
\newblock Mixed precision training.
\newblock {\em arXiv preprint arXiv:1710.03740}, 2017.

\bibitem{narayanan2019pipedream}
Deepak Narayanan, Aaron Harlap, Amar Phanishayee, Vivek Seshadri, Nikhil~R
  Devanur, Gregory~R Ganger, Phillip~B Gibbons, and Matei Zaharia.
\newblock Pipedream: generalized pipeline parallelism for dnn training.
\newblock In {\em Proceedings of the 27th ACM Symposium on Operating Systems
  Principles}, pages 1--15, 2019.

\bibitem{narayanan2021efficient}
Deepak Narayanan, Mohammad Shoeybi, Jared Casper, Patrick LeGresley, Mostofa
  Patwary, Vijay~Anand Korthikanti, Dmitri Vainbrand, Prethvi Kashinkunti,
  Julie Bernauer, Bryan Catanzaro, et~al.
\newblock Efficient large-scale language model training on gpu clusters.
\newblock {\em arXiv preprint arXiv:2104.04473}, 2021.

\bibitem{paszke2019pytorch}
Adam Paszke, Sam Gross, Francisco Massa, Adam Lerer, James Bradbury, Gregory
  Chanan, Trevor Killeen, Zeming Lin, Natalia Gimelshein, Luca Antiga, et~al.
\newblock Pytorch: An imperative style, high-performance deep learning library.
\newblock {\em arXiv preprint arXiv:1912.01703}, 2019.

\bibitem{raffel2019exploring}
Colin Raffel, Noam Shazeer, Adam Roberts, Katherine Lee, Sharan Narang, Michael
  Matena, Yanqi Zhou, Wei Li, and Peter~J Liu.
\newblock Exploring the limits of transfer learning with a unified text-to-text
  transformer.
\newblock {\em arXiv preprint arXiv:1910.10683}, 2019.

\bibitem{rajbhandari2020zero}
Samyam Rajbhandari, Jeff Rasley, Olatunji Ruwase, and Yuxiong He.
\newblock Zero: Memory optimizations toward training trillion parameter models.
\newblock In {\em SC20: International Conference for High Performance
  Computing, Networking, Storage and Analysis}, pages 1--16. IEEE, 2020.

\bibitem{rajbhandari2021zero}
Samyam Rajbhandari, Olatunji Ruwase, Jeff Rasley, Shaden Smith, and Yuxiong He.
\newblock Zero-infinity: Breaking the gpu memory wall for extreme scale deep
  learning.
\newblock {\em arXiv preprint arXiv:2104.07857}, 2021.

\bibitem{rasley2020deepspeed}
Jeff Rasley, Samyam Rajbhandari, Olatunji Ruwase, and Yuxiong He.
\newblock Deepspeed: System optimizations enable training deep learning models
  with over 100 billion parameters.
\newblock In {\em Proceedings of the 26th ACM SIGKDD International Conference
  on Knowledge Discovery \& Data Mining}, pages 3505--3506, 2020.

\bibitem{ren2021zero}
Jie Ren, Samyam Rajbhandari, Reza~Yazdani Aminabadi, Olatunji Ruwase, Shuangyan
  Yang, Minjia Zhang, Dong Li, and Yuxiong He.
\newblock Zero-offload: Democratizing billion-scale model training.
\newblock {\em arXiv preprint arXiv:2101.06840}, 2021.

\bibitem{sergeev2018horovod}
Alexander Sergeev and Mike Del~Balso.
\newblock Horovod: fast and easy distributed deep learning in tensorflow.
\newblock {\em arXiv preprint arXiv:1802.05799}, 2018.

\bibitem{shazeer2018mesh}
Noam Shazeer, Youlong Cheng, Niki Parmar, Dustin Tran, Ashish Vaswani, Penporn
  Koanantakool, Peter Hawkins, HyoukJoong Lee, Mingsheng Hong, Cliff Young,
  et~al.
\newblock Mesh-tensorflow: Deep learning for supercomputers.
\newblock In {\em Advances in Neural Information Processing Systems}, pages
  10414--10423, 2018.

\bibitem{shazeer2018adafactor}
Noam Shazeer and Mitchell Stern.
\newblock Adafactor: Adaptive learning rates with sublinear memory cost.
\newblock In {\em International Conference on Machine Learning}, pages
  4596--4604. PMLR, 2018.

\bibitem{shoeybi2019megatron}
Mohammad Shoeybi, Mostofa Patwary, Raul Puri, Patrick LeGresley, Jared Casper,
  and Bryan Catanzaro.
\newblock Megatron-lm: Training multi-billion parameter language models using
  model parallelism.
\newblock {\em arXiv preprint arXiv:1909.08053}, 2019.

\bibitem{vaswani2017attention}
Ashish Vaswani, Noam Shazeer, Niki Parmar, Jakob Uszkoreit, Llion Jones,
  Aidan~N Gomez, Lukasz Kaiser, and Illia Polosukhin.
\newblock Attention is all you need.
\newblock {\em arXiv preprint arXiv:1706.03762}, 2017.

\bibitem{wang2021pet}
Haojie Wang, Jidong Zhai, Mingyu Gao, Zixuan Ma, Shizhi Tang, Liyan Zheng,
  Yuanzhi Li, Kaiyuan Rong, Yuanyong Chen, and Zhihao Jia.
\newblock {PET}: Optimizing tensor programs with partially equivalent
  transformations and automated corrections.
\newblock In {\em 15th USENIX Symposium on Operating Systems Design and
  Implementation (OSDI 21)}, pages 37--54, 2021.

\bibitem{wang2019tofu}
Minjie Wang, Chien{-}Chin Huang, and Jinyang Li.
\newblock Supporting very large models using automatic dataflow graph
  partitioning.
\newblock In George Candea, Robbert van Renesse, and Christof Fetzer, editors,
  {\em Proceedings of the Fourteenth EuroSys Conference 2019, Dresden, Germany,
  March 25-28, 2019}, pages 26:1--26:17. {ACM}, 2019.

\bibitem{weng2022mlasas}
Qizhen Weng, Wencong Xiao, Yinghao Yu, Wei Wang, Cheng Wang, Jian He, Yong Li,
  Liping Zhang, Wei Lin, and Yu~Ding.
\newblock {MLaaS} in the wild: Workload analysis and scheduling in large-scale
  heterogeneous gpu clusters.
\newblock In {\em 19th {USENIX} Symposium on Networked Systems Design and
  Implementation ({NSDI} 22)}. {USENIX} Association, 2022.

\bibitem{wu2016google}
Yonghui Wu, Mike Schuster, Zhifeng Chen, Quoc~V Le, Mohammad Norouzi, Wolfgang
  Macherey, Maxim Krikun, Yuan Cao, Qin Gao, Klaus Macherey, et~al.
\newblock Google's neural machine translation system: Bridging the gap between
  human and machine translation.
\newblock {\em arXiv preprint arXiv:1609.08144}, 2016.

\bibitem{gandiva2018}
Wencong Xiao, Romil Bhardwaj, Ramachandran Ramjee, Muthian Sivathanu, Nipun
  Kwatra, Zhenhua Han, Pratyush Patel, Xuan Peng, Hanyu Zhao, Quanlu Zhang, Fan
  Yang, and Lidong Zhou.
\newblock Gandiva: Introspective cluster scheduling for deep learning.
\newblock In {\em 13th {USENIX} Symposium on Operating Systems Design and
  Implementation, {OSDI} 2018, Carlsbad, CA, USA, October 8-10, 2018}, pages
  595--610. {USENIX} Association, 2018.

\bibitem{antman2020}
Wencong Xiao, Shiru Ren, Yong Li, Yang Zhang, Pengyang Hou, Zhi Li, Yihui Feng,
  Wei Lin, and Yangqing Jia.
\newblock Antman: Dynamic scaling on {GPU} clusters for deep learning.
\newblock In {\em 14th {USENIX} Symposium on Operating Systems Design and
  Implementation ({OSDI} 20)}, pages 533--548. {USENIX} Association, November
  2020.

\bibitem{xu2021gspmd}
Yuanzhong Xu, HyoukJoong Lee, Dehao Chen, Blake Hechtman, Yanping Huang, Rahul
  Joshi, Maxim Krikun, Dmitry Lepikhin, Andy Ly, Marcello Maggioni, et~al.
\newblock Gspmd: General and scalable parallelization for ml computation
  graphs.
\newblock {\em arXiv preprint arXiv:2105.04663}, 2021.

\bibitem{xue2020mt5}
Linting Xue, Noah Constant, Adam Roberts, Mihir Kale, Rami Al-Rfou, Aditya
  Siddhant, Aditya Barua, and Colin Raffel.
\newblock mt5: A massively multilingual pre-trained text-to-text transformer.
\newblock {\em arXiv preprint arXiv:2010.11934}, 2020.

\bibitem{yang2021exploring}
An~Yang, Junyang Lin, Rui Men, Chang Zhou, Le~Jiang, Xianyan Jia, Ang Wang, Jie
  Zhang, Jiamang Wang, Yong Li, et~al.
\newblock Exploring sparse expert models and beyond.
\newblock {\em arXiv preprint arXiv:2105.15082}, 2021.

\bibitem{yang2021pipemare}
Bowen Yang, Jian Zhang, Jonathan Li, Christopher R{\'e}, Christopher Aberger,
  and Christopher De~Sa.
\newblock Pipemare: Asynchronous pipeline parallel dnn training.
\newblock {\em Proceedings of Machine Learning and Systems}, 3, 2021.

\bibitem{zheng2020ansor}
Lianmin Zheng, Chengfan Jia, Minmin Sun, Zhao Wu, Cody~Hao Yu, Ameer Haj-Ali,
  Yida Wang, Jun Yang, Danyang Zhuo, Koushik Sen, et~al.
\newblock Ansor: Generating high-performance tensor programs for deep learning.
\newblock In {\em 14th {USENIX} Symposium on Operating Systems Design and
  Implementation ({OSDI} 20)}, pages 863--879, 2020.

\bibitem{zheng2022astitch}
Zhen Zheng, Xuanda Yang, Pengzhan Zhao, Guoping Long, Kai Zhu, Feiwen Zhu,
  Wenyi Zhao, Xiaoyong Liu, Jun Yang, Jidong Zhai, Shuaiwen~Leon Song, and Wei
  Lin.
\newblock Astitch: Enabling a new multi-dimensional optimization space for
  memory-intensive ml training and inference on modern simt architectures.
\newblock In {\em Proceedings of the 27th ACM International Conferenceon
  Architectural Support for Programming Languages and Operating Systems}. ACM,
  2022.

\end{thebibliography}

\end{document}